\documentclass[preprint,12pt,sort&compress]{elsarticle}

\usepackage{graphicx}

\journal{Nuclear Physics A}

\begin{document}

\begin{frontmatter}



\title{Quantum design using a multiple internal reflections method in a study of fusion processes
in the capture of alpha-particles by nuclei}


\author[imp,kinr]{Sergei~P.~Maydanyuk}
\ead{maidan@kinr.kiev.ua}

\author[imp]{Peng-Ming~Zhang}
\ead{zhpm@impcas.ac.cn}

\author[kinr]{Sergei~V.~Belchikov}
\ead{sbelchik@kinr.kiev.ua}

\address[imp]{Institute of Modern Physics, Chinese Academy of Sciences, Lanzhou, 730000, China}
\address[kinr]{Institute for Nuclear Research, National Academy of Sciences of Ukraine, Kiev, 03680, Ukraine}



\begin{abstract}
A high precision method 
to determine fusion in the capture of $\alpha$-particles by nuclei is presented.
For $\alpha$-capture by $^{40}{\rm Ca}$ and $^{44}{\rm Ca}$, such an approach gives 
(1) the parameters of the $\alpha$--nucleus potential and
(2) fusion probabilities.
This method found new parametrization and fusion probabilities and decreased the error 
by $41.72$ times for $\alpha + ^{40}{\rm Ca}$ and $34.06$ times for $\alpha + ^{44}{\rm Ca}$ in a description of experimental data
in comparison with existing results.
We show that
the sharp angular momentum cutoff 
proposed by Glas and Mosel is a rough approximation,
Wong's formula and the Hill-Wheeler approach determine the penetrability of the barrier without a correct consideration of the barrier shape,
and the WKB approach gives reduced fusion probabilities.
%
%
%
%
%
Based on our fusion probability formula, we explain the difference between experimental cross-sections
for $\alpha + ^{40}{\rm Ca}$ and $\alpha + ^{44}{\rm Ca}$,
which is connected with the theory of coexistence of the spherical and deformed shapes in the ground state for nuclei near the neutron magic shell $N=20$.
To provide deeper insight into the physics of nuclei with the new magic number $N=26$,
the cross-section for $\alpha + ^{46}{\rm Ca}$ is predicted for future experimental tests.
The role of nuclear deformations in calculations of the fusion probabilities is analyzed.
\end{abstract}



\end{frontmatter}


\section{Introduction
\label{introduction}}

Understanding fusion in nuclear reactions is a fundamental problem of physics \cite{Balantekin.1998.RMP}.
A topic where such processes are important is the synthesis of superheavy nuclei \cite{Hofmann.2000.RMP,Sobiczewski.2007.PPNP}.
Here, fusion has a crucial role, but its analysis is complicated.
Deep insight into the physics of fusion can be obtained from the capture of the $\alpha$-particles by the nuclei.

Information about fusion in $\alpha$-capture forms our understanding of interactions between the $\alpha$-particles and nuclei at distances, where an important role is attributed to the formation of the nucleus from two colliding nuclear objects inside the spatial nuclear region.
$\alpha$-nucleus interactions have been studied most intensively
in the context of the $\alpha$-decay of nuclei
(see experimental information~%
\cite{Buck.1991.JPG,Buck.1992.PRC,Buck.1993.ADNDT,%
Duarte.2002.ADNDT,Dasgupta-Schubert.2007.ADNDT,Firestone.1996.book,%
Akovali.1998.NDS,Audi.2003.NPA,Gupta.2005.NDS,Belli.2007.NPA,Nishio.2003.PRC,Karamian.2007.PRC,Zhang.2014.PRC},
various microscopic models~\cite{Lovas.1998.PRep,Tomas.1954.PTP,%
Kadmenskii.1976.SJPN,Kadmenskii.1972.SJPN,Kadmenskii.2007.PPN,%
Stewart.1996.NPA,Delion.1992.PRC,Delion.1994.PRC,Delion.2003.PRC,Delion.2013.PRC,%
Id_Betan.2012.PRC,Silesteanu.2001.NPA,Silesteanu.2010.RJP,Silesteanu.2012.ADNDT},
macroscopic cluster models~%
\cite{Strutinsky.1955.RSAS,Strutinsky.1957.RSAS,Royer.2000.JPG,Moustabchir.2001.NPA,Basu.2003.PLB,%
Denisov.2005.PHRVA,Denisov.2009.PRC.v79,Denisov.2009.PRC.v80,Denisov.2009.ADNDT,%
Xu.2006.PRC,Medeiros.2006.PRC,Samanta.2007.NPA,Bhagwat.2008.JPG,Zhang.2008.PRC},
fission models~\cite{Poenaru.2006.PRC,Sobiczewski.2007.PPNP}),
and the scattering of the $\alpha$-particles off
nuclei~\cite{Huizenga.1962.NP,Nolte.1987.PRC,Atzrott.1996.PRC,Demetriou.2002.NPA,Avrigeanu.1996.PRC}.
The physics of the fusion processes during $\alpha$-capture have been investigated less deeply~\cite{Denisov.2005.PHRVA,Denisov.2009.PRC.v79,Denisov.2009.PRC.v80,Denisov.2009.ADNDT}.
Evaluations of the $\alpha$-particle capture rates show that they play an important role in nuclear reactions in
stars~\cite{Mohr.2000.PRC,Demetriou.2002.NPA,Rauscher.2000.NPA}.

A prevailing approach for the determination of the capture cross-sections of the $\alpha$-particle by the nucleus is based on calculations of the penetrability of the potential barriers, where information about the fusion is not included. While many approaches have been developed for the calculation of the penetrability,
there is no generally accepted method to estimate the fusion.
Experimentally, these reactions have not been studied deeply: we have the cross-sections for capture of the $\alpha$-particles by nuclei $^{40}{\rm Ca}$, $^{44}{\rm Ca}$ \cite{Eberhard.1979.PRL}, $^{59}{\rm Co}$ \cite{DAuria.1968.PR}, $^{208}{\rm Pb}$~\cite{Barnett.2000.PRC}, and $^{209}{\rm Be}$~\cite{Barnett.2000.PRC}.

We shall be interesting in
information about fusion that can be extracted from the available experimental data.
Thus, the aim of this paper is to construct a new approach for obtaining such information.
The idea of such an approach can be similar to the inverse theory of scattering~\cite{Zakhariev.1985.book}.
Eberhard et al 
proposed a relation that gives information about fusion in the $\alpha$-capture and
compares calculated cross-sections with experimental data
at selected energies \cite{Eberhard.1979.PRL}.
%
Analyzing the existed experimental data, we observe an interesting difference in fusion between $\alpha$-captures by the $^{40}{\rm Ca}$ and $^{44}{\rm Ca}$ nuclei, while the inclusion of the other nuclei into the analysis does not result in principally new questions.

A crucial point in such a task is the determination of the barrier penetrability.
It turns out that even small improvements in the Wentzel-Kramers-Brillouin (WKB) formula of the penetrability require the reconsideration of the connection between boundary conditions and even the inclusion of the initial (or final) condition into the stationary picture of the studied process.
After such a modification, the penetrability becomes more sensitive to the shape of the barrier, and it is already dependent on the potential outside the tunneling region.
A fully quantum realization of the approach
leads to the essential influence of the calculated penetrability on additional parameters (see \cite{Maydanyuk.2011.JMP} for details, proofs, demonstrations).
The role of such additional parameters in the determination of the penetrability is larger in some orders in comparison with changes of the resulting probability as a result of any variations in the nuclear deformations and parameters of the interacting potentials if the penetrability is not calculated in the fully quantum approach.
In this paper, we shall therefore restrict ourselves to the spherically symmetric $\alpha$-nucleus potentials.
However, the role of the nuclear deformations in the determination of the fusion probabilities in the $\alpha$-capture task
is analyzed in~\ref{sec.app.6}.

We generalize the method of multiple internal reflections
(MIR method, see~\cite{Maydanyuk.2000.UPJ,Maydanyuk.2002.JPS,Maydanyuk.2002.PAST,Maydanyuk.2003.PhD-thesis,%
Maydanyuk.2006.FPL,Maydanyuk.2011.JMP}) to describe the capture of the $\alpha$-particles by nuclei.
We analyze the \emph{sharp angular momentum cutoff approach} previously introduced by Glas and Mosel in~\cite{Glas.1975.NPA, Glas.1974.PRC}, which has been widely used (for example, see \cite{Aguilera.2012.PRC,Kumari.2013.NPA}). We analyze the WKB approach, Wong's formula~\cite{Wong.1973.PRL} and the Hill-Wheeler approach~\cite{Hill_Wheeler.1953.PR} in calculations of the cross-sections.

We focus on understanding reasons for difference between the cross-sections for $\alpha + ^{40}{\rm Ca}$ and $\alpha + ^{44}{\rm Ca}$.
The inclusion of the fusion process into the $\alpha$-capture task allows us to essentially improve the agreement between the calculations and experimental data,
and it provides a deeper picture of the $\alpha$-capture than the sharp angular momentum cut off approach.
We show that the MIR approach, in contrast to other approaches, satisfies all tests of quantum mechanics, resolves the problem of indefinite additional parameters described above (fixing the result and excluding errors),
and it is the most effective and accurate method for studying fusion in this task.
We find a new formula for fusion probability and propose $\alpha$-capture estimations for the $^{46}{\rm Ca}$ nucleus with the new neutron magic number $N=26$ \cite{Penionzhkevich.2006.EPAN}.


\section{Method
\label{sec.2}}

\subsection{Cross-section of $\alpha$-capture and sharp angular momentum cut-off
\label{sec.2.1}}

The cross-section of capture that includes the fusion of the $\alpha$-particles with nuclei can be defined
as~\cite{Eberhard.1979.PRL}
\begin{equation}
  \sigma_{\rm capture} (E) =
  \displaystyle\frac{\pi\, \hbar^{2}}{2\,m\,\tilde{E}}\;
    \displaystyle\sum\limits_{l=0}^{+\infty}
    (2l+1)\, T_{l}\, P_{l},
\label{eq.2.1.1}
\end{equation}
where
$E$ is the kinetic energy of the $\alpha$-particle incident on the nucleus in the laboratory frame,
$\tilde{E}$ is the kinetic energy of the relative motion of the $\alpha$-particle of the nucleus in the center-of-mass frame (we use $E \simeq \tilde{E}$),
$m$ is the reduced mass of the $\alpha$-particle and nucleus,
$P_{l}$ is the probability for fusion of the $\alpha$-particle and nucleus, and
$T_{l}$ is the penetrability of the barrier.
In the WKB approach, this coefficient is defined as
\begin{equation}
  T_{WKB}(\tilde{E}) =
  \exp\;
  \Biggl\{
    -2 \displaystyle\int\limits_{R_{2}}^{R_{3}}
    \sqrt{\displaystyle\frac{2m}{\hbar^{2}}\: \Bigl(\tilde{E} - V(r)\Bigr)} \; dr
  \Biggr\},
\label{eq.2.1.2}
\end{equation}
where $R_{2}$ and $R_{3}$ are the second and third turning points determining the tunneling region.%
\footnote{The first turning point is defined as the intersection of the potential and energy near zero, where the centrifugal component is the main contribution}.

We shall consider the capture when the fragment has passed to the internal potential well
after tunneling through the barrier.
Solving a set of classical motion equations, one can find the critical angular momentum $l_{\rm cr}$ when all trajectories at $l<l_{\rm cr}$ lead to capture.
Glas and Mosel introduced the sharp angular momentum cutoff~\cite{Glas.1974.PRC,Glas.1975.NPA}, where
\begin{equation}
  P_{l} =
  \left\{
  \begin{array}{lll}
    1 & \mbox{\rm at } & l \le l_{\rm cr}, \\
    0 & \mbox{\rm at } & l > l_{\rm cr}.
  \end{array}
\right.
\label{eq.2.1.3}
\end{equation}
%
In some papers (for example, see~\cite{Eberhard.1979.PRL}), an assumption that $T_{l}=1$ for $l<l_{\rm gr}$ is imposed (where $l_{\rm gr} > l_{\rm cr}$),
which transforms formula (\ref{eq.2.1.2}) into
\begin{equation}
  \sigma_{\rm capture} (E) =
  \displaystyle\frac{\pi\, \hbar^{2}}{2\,m\,\tilde{E}}\;
  (2l_{\rm cr}+1)^{2}.
\label{eq.2.1.4}
\end{equation}
The critical momenta $l_{\rm cr}$ are found after comparison of the cross-sections obtained by this formula with experimental data. Eberhard et al in~\cite{Eberhard.1979.PRL} obtained $l_{\rm cr} = 9.9$ for $\alpha + ^{40}{\rm Ca}$ and $l_{\rm cr} = 10.9$ for $\alpha + ^{44}{\rm Ca}$.
We shall reconsider this process for the quantum consideration that leads to a set of tunneling transitions
at each allowed orbital momentum $l$ according to formula (\ref{eq.2.1.1}).


\subsection{Multiple internal reflections method in the barrier penetrability determination of an arbitrary shape
\label{sec.2.2}}

We shall study the capture of the $\alpha$-particle by the nucleus in a spherically symmetric scenario
(the role of nuclear deformations in the determination of the fusion probabilities in the $\alpha$-capture task
is analyzed in~\ref{sec.app.6}).
To apply the idea of multiple internal reflections to study packet tunneling through complicated realistic barriers, let us consider the radial barrier of an arbitrary shape, which has successfully been
approximated by a sufficiently large number $N$ of rectangular steps:
\begin{equation}
  V(r) = \left\{
  \begin{array}{cll}
    V_{1}   & \mbox{at } r_{\rm min} < r \leq r_{1}      & \mbox{(region 1)}, \\
    V_{2}   & \mbox{at } r_{1} \leq r \leq r_{2}         & \mbox{(region 2)}, \\
    \ldots   & \ldots & \ldots \\
    V_{N}   & \mbox{at } r_{N-1} \leq r \leq r_{\rm max} & \mbox{(region $N$)},
  \end{array} \right.
\label{eq.2.2.1}
\end{equation}
where $V_{j}$ are constants ($j = 1 \ldots N$)%
\footnote{We used this approximation for barriers of the proton and $\alpha$-decay for the number of nuclei, where the width of each step is equal to 0.01~fm, and we demonstrated the stability of the calculations of all amplitudes of the wave function and penetrability~\cite{Maydanyuk.2011.JMP}. We thus have effective tools for a detailed study of the quantum processes of tunneling and penetrability.}.
Let us denote the first region with a left boundary at point $r_{\rm min}$ (we denote it also as $r_{\rm capture}$), and we shall assume that the capture of the $\alpha$-particle by the nucleus in this region occurs after its tunneling through the barrier. We shall be interested in solutions for the above barrier energies, while the solution for tunneling could be subsequently obtained by changing $i\,\xi_{i} \to k_{i}$.
A general solution of the wave function (up to
its normalization) has the following form:
\begin{equation}
  \psi(r, \theta, \varphi) =  \frac{\chi(r)}{r} Y_{lm}(\theta, \varphi),
\label{eq.2.2.2}
\end{equation}
\begin{equation}
\chi(r) = \left\{
\begin{array}{lll}
   \alpha_{1}\, e^{ik_{1}r} + \beta_{1}\, e^{-ik_{1}r},
     & \mbox{at } r_{\rm min} < r \leq r_{1} & \mbox{(region 1)}, \\
   \alpha_{2}\, e^{ik_{2}r} + \beta_{2}\, e^{-ik_{2}r},
     & \mbox{at } r_{1} \leq r \leq r_{2} & \mbox{(region 2)}, \\
   \ldots & \ldots & \ldots \\
   \alpha_{N-1}\, e^{ik_{N-1}r} + \beta_{N-1}\, e^{-ik_{N-1}r}, &
     \mbox{at } r_{N-2} \leq r \leq r_{N-1} & \mbox{(region $N-1$)}, \\
   e^{-ik_{N}r} + A_{R}\,e^{ik_{N}r}, & \mbox{at } r_{N-1} \leq r \leq r_{\rm max} & \mbox{(region $N$)},
\end{array} \right.
\label{eq.2.2.3}
\end{equation}
where $\alpha_{j}$ and $\beta_{j}$ are unknown amplitudes, $A_{T}$ and $A_{R}$ are unknown amplitudes of transition and reflection, $Y_{lm}(\theta ,\varphi)$ is the spherical function, and $k_{j} = \frac{1}{\hbar}\sqrt{2m(\tilde{E}-V_{j})}$ are complex wave numbers. We have fixed the normalization so that the modulus of amplitude of the starting wave $e^{-ik_{N}r}$ equals unity. We shall search a solution to this problem by the multiple internal reflections approach.

According to the multiple internal reflections method, the scattering of a particle on the barrier is sequentially considered by steps of propagation of the wave packet relative to each boundary of the barrier (the idea of this approach can be understood most clearly in the problem of tunneling through the simplest rectangular barrier, see~\cite{Maydanyuk.2002.JPS, Maydanyuk.2003.PhD-thesis, Maydanyuk.2006.FPL}, where one can find proof of this fully quantum exactly solvable method and analyze its properties). Each step in such a consideration of packet propagation is similar to one of the first independent $2N-1$ steps. From the analysis of these steps, we find recurrent relations for the calculation of the unknown amplitudes $A_{T}^{(n)}$, $A_{R}^{(n)}$, $\alpha_{j}^{(n)}$ and $\beta_{j}^{(n)}$ for an arbitrary step with number $n$ (here, index $j$ corresponds to the number of region $V_{j}$, and the logic of the definition of these amplitudes can be found in Appendix~A in~\cite{Maydanyuk.2011.JMP}). In the summation of these relations from each step, we impose the continuity condition for the full (summarized) wave function and its derivative relative to the corresponding boundary.

According to the analysis of waves propagating in a region with an arbitrary number $j$ on the arbitrary step, each wave can be represented as a multiplication of the exponential factor $e^{\pm i\,k_{j}\,r}$ and constant amplitude. In practical calculations, the difficulty consists in the determination of these unknown amplitudes.
However, to make such calculations as easy as possible for an arbitrary step, one can rewrite the amplitude of the wave that has transmitted through the boundary with number $j$ as the product of the amplitude of the corresponding wave incident on this boundary and the new factor $T_{j}^{\pm}$ (i.~e., the amplitude of transition through the boundary with number $j$).
The bottom index indicates the number of the boundary, and the upper sign ``$+$'' or ``$-$'' is the direction of the incident wave to the right or left, respectively.
We associate the amplitude of the reflected wave from the boundary with number $j$ with the amplitude of the wave incident on this boundary via new factors $R_{j}^{\pm}$.
The coefficients $T_{1}^{\pm}$, $T_{2}^{\pm}$, $T_{3}^{\pm}$\ldots and $R_{1}^{\pm}$, $R_{2}^{\pm}$, $R_{3}^{\pm}$\ldots can be found from the recurrence relations shown above~\cite{Maydanyuk.2011.JMP}.
We calculate
$T_{1}^{\pm}$, $T_{2}^{\pm}$ \ldots $T_{N-1}^{\pm}$ and $R_{1}^{\pm}$,
$R_{2}^{\pm}$ \ldots $R_{N-1}^{\pm}$ as
\begin{equation}
\begin{array}{ll}
\vspace{2mm}
   T_{j}^{+} = \displaystyle\frac{2k_{j}}{k_{j}+k_{j+1}} \,e^{i(k_{j}-k_{j+1}) r_{j}}, &
   T_{j}^{-} = \displaystyle\frac{2k_{j+1}}{k_{j}+k_{j+1}} \,e^{i(k_{j}-k_{j+1}) r_{j}}, \\
   R_{j}^{+} = \displaystyle\frac{k_{j}-k_{j+1}}{k_{j}+k_{j+1}} \,e^{2ik_{j}r_{j}}, &
   R_{j}^{-} = \displaystyle\frac{k_{j+1}-k_{j}}{k_{j}+k_{j+1}} \,e^{-2ik_{j+1}r_{j}}.
\end{array}
\label{eq.2.2.4}
\end{equation}

Now, we consider the wave propagating in region with number $j$, which is incident from the right on the potential barrier with the right boundary at point $r_{j-1}$ (and left boundary at point $r_{1}$). Let us find the wave reflected from this complicated barrier.
This wave should combine all waves formed as a result of multiple internal reflections and propagations relative boundaries $r_{1}$ \ldots $r_{j-1}$ and leave such a barrier.
We define the reflection amplitude $\tilde{R}_{j-1}^{+}$ of such a summarized wave as
\begin{equation}
\begin{array}{l}
   \vspace{1mm}
   \tilde{R}_{j-1}^{+} =
     R_{j-1}^{+} + T_{j-1}^{+} \tilde{R}_{j}^{+} T_{j-1}^{-}
     \Bigl(1 + \sum\limits_{m=1}^{+\infty} (\tilde{R}_{j}^{+}R_{j-1}^{-})^{m} \Bigr) =
     R_{j-1}^{+} +
     \displaystyle\frac{T_{j-1}^{+} \tilde{R}_{j}^{+} T_{j-1}^{-}} {1 - \tilde{R}_{j}^{+} R_{j-1}^{-}}.
\end{array}
\label{eq.2.2.5}
\end{equation}
Correspondingly, we also define
\begin{equation}
\begin{array}{lcl}
%
  \vspace{1mm}
  \tilde{R}_{j+1}^{-} & = &
    R_{j+1}^{-} + T_{j+1}^{-} \tilde{R}_{j}^{-} T_{j+1}^{+}
    \Bigl(1 + \sum\limits_{m=1}^{+\infty} (R_{j+1}^{+} \tilde{R}_{j}^{-})^{m} \Bigr) = 
    R_{j+1}^{-} +
    \displaystyle\frac{T_{j+1}^{-} \tilde{R}_{j}^{-} T_{j+1}^{+}} {1 - R_{j+1}^{+} \tilde{R}_{j}^{-}}, \\

  \vspace{1mm}
  \tilde{T}_{j+1}^{+} & = &
    \tilde{T}_{j}^{+} T_{j+1}^{+}
    \Bigl(1 + \sum\limits_{m=1}^{+\infty} (R_{j+1}^{+} \tilde{R}_{j}^{-})^{m} \Bigr) =
    \displaystyle\frac{\tilde{T}_{j}^{+} T_{j+1}^{+}} {1 - R_{j+1}^{+} \tilde{R}_{j}^{-}}, \\

  \tilde{T}_{j-1}^{-} & = &
    \tilde{T}_{j}^{-} T_{j-1}^{-}
    \Bigl(1 + \sum\limits_{m=1}^{+\infty} (R_{j-1}^{-} \tilde{R}_{j}^{+})^{m} \Bigr) =
    \displaystyle\frac{\tilde{T}_{j}^{-} T_{j-1}^{-}} {1 - R_{j-1}^{-} \tilde{R}_{j}^{+}}.
\end{array}
\label{eq.2.2.6}
\end{equation}
In such a summation, we have recurrent relations, which connect all amplitudes.
We now choose the following values
\begin{equation}
\begin{array}{cccc}
  \tilde{R}_{N-1}^{+} = R_{N-1}^{+}, & \quad
  \tilde{R}_{1}^{-} = R_{1}^{-}, & \quad
  \tilde{T}_{1}^{+} = T_{1}^{+}, & \quad
  \tilde{T}_{N-1}^{-} = T_{N-1}^{-},
\end{array}
\label{eq.2.2.7}
\end{equation}
as our starting point and consequently calculate all amplitudes $\tilde{R}_{N-2}^{+}$ \ldots $\tilde{R}_{1}^{+}$, $\tilde{R}_{2}^{-}$ \ldots $\tilde{R}_{N-1}^{-}$ and $\tilde{T}_{2}^{+}$ \ldots $\tilde{T}_{N-1}^{+}$.
We find the coefficients $\alpha_{j}$ and $\beta_{j}$:
\begin{equation}
\begin{array}{lcl}
   \vspace{1mm}
   \alpha_{j} & = &
   \sum\limits_{n=1}^{+\infty} \alpha_{j}^{(n)} =
     \tilde{T}_{j-1}^{+}
     \Bigl(1 + \sum\limits_{m=1}^{+\infty} (R_{j}^{+} \tilde{R}_{j-1}^{-})^{m} \Bigr) = 
     \displaystyle\frac{\tilde{T}_{j-1}^{+}} {1 - R_{j}^{+} \tilde{R}_{j-1}^{-}} =
     \displaystyle\frac{\tilde{T}_{j}^{+}} {T_{j}^{+}}, \\

   \beta_{j} & = &
   \sum\limits_{n=1}^{+\infty} \beta_{j}^{(n)} =
     \tilde{T}_{j-1}^{+}
     \Bigl(1 + \sum\limits_{m=1}^{+\infty} (R_{j}^{+} \tilde{R}_{j-1}^{-})^{m} \Bigr)\, R_{j}^{+} = 
     \displaystyle\frac{\tilde{T}_{j-1}^{+}\, R_{j}^{+}} {1 - R_{j}^{+} \tilde{R}_{j-1}^{-}} =
     \displaystyle\frac{\tilde{T}_{j}^{+} R_{j}^{+}} {T_{j}^{+}} =
     \alpha_{j} \cdot R_{j}^{+},
\end{array}
\label{eq.2.2.8}
\end{equation}
the amplitudes of transition and reflection concerning the whole barrier:
\begin{equation}
\begin{array}{ll}
  A_{T} = \sum\limits_{n=1}^{+\infty} A_{T}^{(n)} = \tilde{T}_{1}^{+}, &
  A_{R} = \sum\limits_{n=1}^{+\infty} A_{R}^{(n)} = \tilde{R}_{N-1}^{+}
\end{array}
\label{eq.2.2.9}
\end{equation}
and the corresponding coefficients of penetrability $T_{MIR}$%
\footnote{We shall analyze the penetrabilities and cross-sections obtained by the multiple internal reflections method and WKB method. To distinguish the calculated results, we shall add abbreviation \emph{``MIR''} to the \emph{``multiple internal reflections method''} and \emph{WKB} to the WKB approach.}
and reflection $R_{MIR}$:
\begin{equation}
\begin{array}{cc}
  T_{MIR} \equiv \displaystyle\frac{k_{N}}{k_{1}}\; \bigl|A_{T}\bigr|^{2}, &
  R_{MIR} \equiv \bigl|A_{R}\bigr|^{2}.
\end{array}
\label{eq.2.2.10}
\end{equation}
We check the property
\begin{equation}
\begin{array}{ccc}
  \displaystyle\frac{k_{N}}{k_{1}}\; |A_{T}|^{2} + |A_{R}|^{2} = 1 & \mbox{ or }&
  T_{MIR} + R_{MIR} = 1,
\end{array}
\label{eq.2.2.11}
\end{equation}
which indicates whether the MIR method gives the proper solutions for the wave function. If the energy of the particle is located below the height of step with number $m$, then for a description of the transition of the particle through the barrier with its
tunneling, the following change must be performed:
\begin{equation}
  k_{m} \to i\,\xi_{m},
\label{eq.2.2.12}
\end{equation}
where $\xi_{m} = \frac{1}{\hbar}\sqrt{2m(V_{m}-E)}$. For two rectangular barrier steps of arbitrary sizes we obtain coincident amplitudes of the wave function calculated by MIR approach, with corresponding amplitudes obtained in the standard approach of quantum mechanics (up to the first 15 digits). Increasing the number of steps up into the thousands maintains such an accuracy in coincidence between calculations in the MIR approach and the standard approach of quantum mechanics, and condition (\ref{eq.2.2.11}) is fulfilled (see \ref{sec.app.1}, where we briefly present algorithms for the standard approach of quantum mechanics). This is an important test confirming the reliability of the MIR method.
We have therefore obtained the full coincidence between the solutions for all amplitudes obtained by the MIR approach and the standard approach of quantum mechanics.
A comparative analysis of the accuracy of this method and another method in determining
the penetrability and reflection coefficients on the basis of wave functions obtained via the direct integration of the radial Schr\"{o}dinger equation with high accuracy using the Numerov technique
is given in~\ref{sec.app.5}.


\subsection{$\alpha$-nucleus potential and minimization method
\label{sec.2.3}}

We define interactions between the $\alpha$-particle and nucleus by the potential
\begin{equation}
  V (r, l, Q) = v_{C} (r) + v_{N} (r, Q) + v_{l} (r),
\label{eq.2.3.1}
\end{equation}
where $v_{C} (r)$, $v_{N} (r, Q)$ and $v_{l} (r)$ are the Coulomb, nuclear and centrifugal components.
In the spherically symmetric consideration of the $\alpha$-capture, we have
\begin{equation}
\begin{array}{ll}
  v_{C} (r) =
  \left\{
  \begin{array}{ll}
    \displaystyle\frac{2 Z e^{2}} {r} &
      \mbox{at  } r \ge r_{\rm m}, \\
    \displaystyle\frac{Z e^{2}} {r_{\rm m}}\;
      \biggl\{ 3 -  \displaystyle\frac{r^{2}}{r_{\rm m}^{2}} \biggr\} &
      \mbox{at  } r < r_{\rm m},
  \end{array}
  \right. &
  v_{N} (r, Q) = \displaystyle\frac{V(A,Z,Q)} {1 + \exp{\displaystyle\frac{r-r_{\rm m}} {d}}}
\end{array}
\label{eq.2.3.2}
\end{equation}
%
%
and
\begin{equation}
  r_{\rm m} = 1.5268 + R.
\label{eq.2.3.4}
\end{equation}
We have previously used parametrization~\cite{Denisov.2005.PHRVA},
with a special orientation toward the description of $\alpha$-capture.
%
%
%
%
%
%
%
$V(A,Z,Q)$ is the strength of the nuclear component,
$A$ and $Z$ are the numbers of nucleons and protons in the nucleus,
$Q$ is the $Q$-value of the $\alpha$-capture,
$r_{\rm m}$ is the effective radius of the nuclear component,
$R$ is the radius of the nucleus, and
$d$ is the diffuseness ($R$, $r_{\rm m}$, $d$ are in fm).
We define the following functions of errors:
\begin{equation}
\begin{array}{lcl}
  \varepsilon_{1} & = &
  \displaystyle\frac{1}{N}
  \displaystyle\sum\limits_{k=1}^{N}
    \displaystyle\frac{\Bigl|\sigma^{\rm (theor)} (E_{k}) - \sigma^{\rm (exp)} (E_{k}) \Bigr|}{\sigma^{\rm (exp)} (E_{N})}, \\
  \varepsilon_{2} & = &
  \displaystyle\frac{1}{N}
  \displaystyle\sum\limits_{k=1}^{N}
    \displaystyle\frac{\Bigl|\sigma^{\rm (theor)} (E_{k}) - \sigma^{\rm (exp)} (E_{k}) \Bigr|}{\sigma^{\rm (exp)} (E_{k})}, \\
  \varepsilon_{3} & = &
  \displaystyle\frac{1}{N}
  \displaystyle\sum\limits_{k=1}^{N}
    \displaystyle\frac{\Bigl|\ln(\sigma^{\rm (theor)} (E_{k})) - \ln(\sigma^{\rm (exp)} (E_{k})) \Bigr|}{\ln(\sigma^{\rm (exp)} (E_{k}))},
\end{array}
\label{eq.2.3.11}
\end{equation}
where $\sigma^{\rm (theor)} (E_{k})$ and $\sigma^{\rm (exp)} (E_{k})$ are theoretical and experimental cross-sections of capture at energy $E_{k}$, and the summation is performed over experimental data.
We shall search the fusion probabilities $p_{0}$ \ldots $p_{\rm max}$ for the minimal characteristic (\ref{eq.2.3.11}).
We shall call such an approach the minimization method.

\section{Analysis
\label{sec.3}}

\subsection{Dependence of the penetrability on the spatial localization of capture and orbital momenta
\label{sec.3.1}}

\begin{figure}[htbp]
%
%
\hspace{-5mm}
\includegraphics[width=76mm]{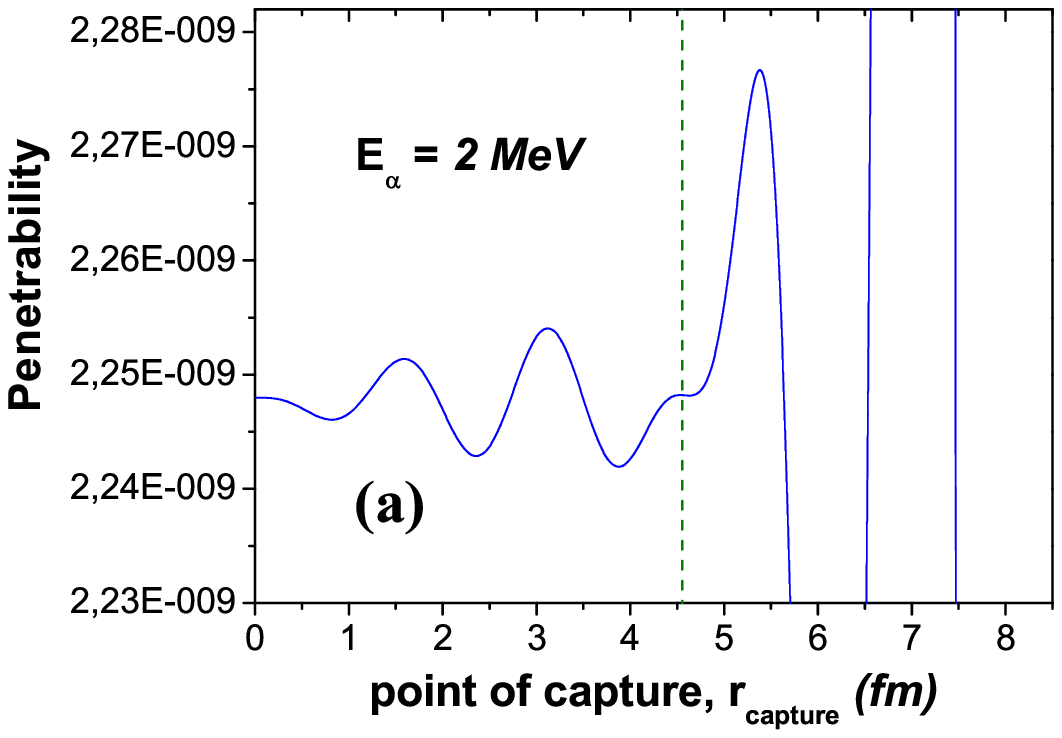}
\hspace{-7mm}\includegraphics[width=76mm]{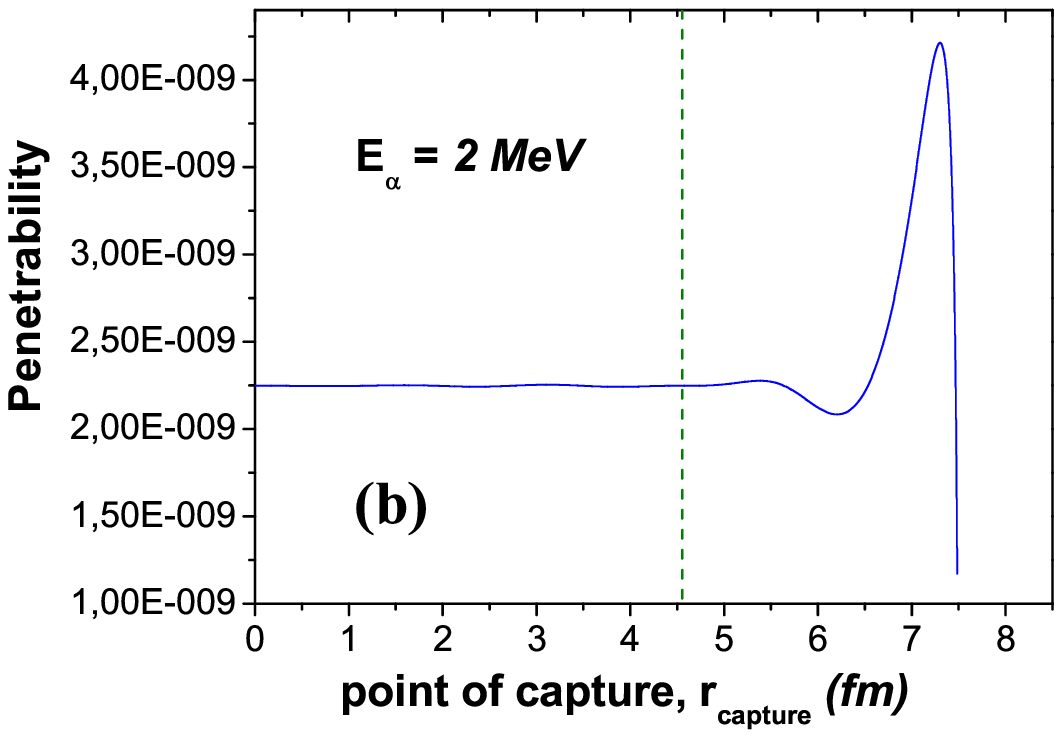}
\vspace{-5mm}
\caption{\small 
The penetrability of the barrier depends on the position of the capture localization of the $\alpha$-particle $r_{\rm capture}$ for the reaction of $\alpha + ^{44}{\rm Ca}$ at $l=0$ and the energy of the $\alpha$-particle $E_{\alpha}=2$~MeV
(parameters of calculations: 10000 intervals at $r_{\rm max}=70$~fm,
parametrization~\cite{Denisov.2005.PHRVA}).
\label{fig.3.0.1}}
\end{figure}

For the analysis of the method and general properties of $\alpha$-capture, we shall choose the $^{44}{\rm Ca}$ nucleus. Let us consider how the position of $r_{\rm capture}$ influences the barrier penetrability (as we assume capture occurs when the maximum of the wave packet describing the $\alpha$-particle is located within region $M$).
The results of such calculations for the energy 2~MeV of the $\alpha$-particle are presented in Fig.~\ref{fig.3.0.1}.
One can see that small variations in the position of $r_{\rm capture}$ lead to an essential change of the calculated penetrability (for example, the influence of the position of $r_{\rm capture}$ is essentially higher than that of the deformation of the nucleus; therefore, there is no sense in studying the role of nuclear deformations without a previous consideration of this influence). The importance of the proper choice of this parameter in calculations of the capture cross-sections is now becoming clear. This dependence has an oscillating behavior that is characterized by the manifestation of the wave nature (origin) of the tunneling processes, and it was previously observed by us in the scenarios of proton and $\alpha$-decays \cite{Maydanyuk.2011.JMP}.

As a possible value for $r_{\rm capture}$, we choose the coordinate corresponding to the minimal amplitude variation of the penetrability oscillations, which corresponds to the highest stability of the penetrability for possible perturbations (we do not consider similar behavior of the penetrability at $r \to 0$). Our performed analysis has shown that such a point is the same for different energies of the $\alpha$-particle, and it corresponds to the minimum of the internal potential well (we obtain $r_{\rm capture} = 0.44$~fm).
We shall choose such a definition of $r_{\rm capture}$ for further calculations.

\begin{figure}[htbp]
\centerline{%
\includegraphics[width=78mm]{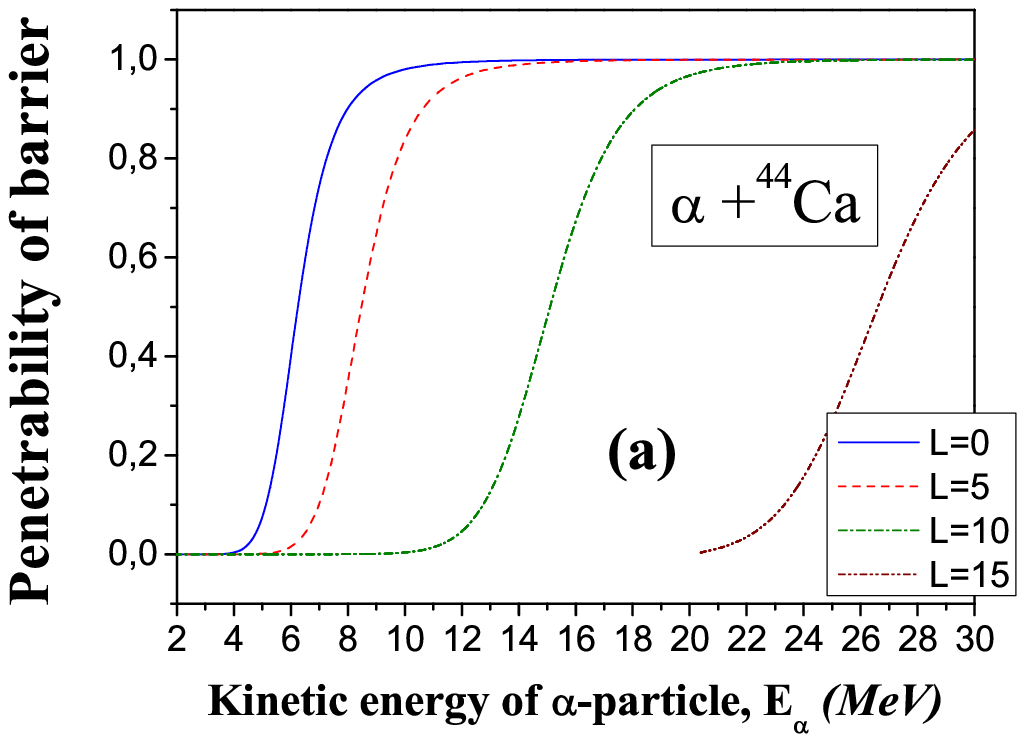}
\hspace{-5mm}\includegraphics[width=78mm]{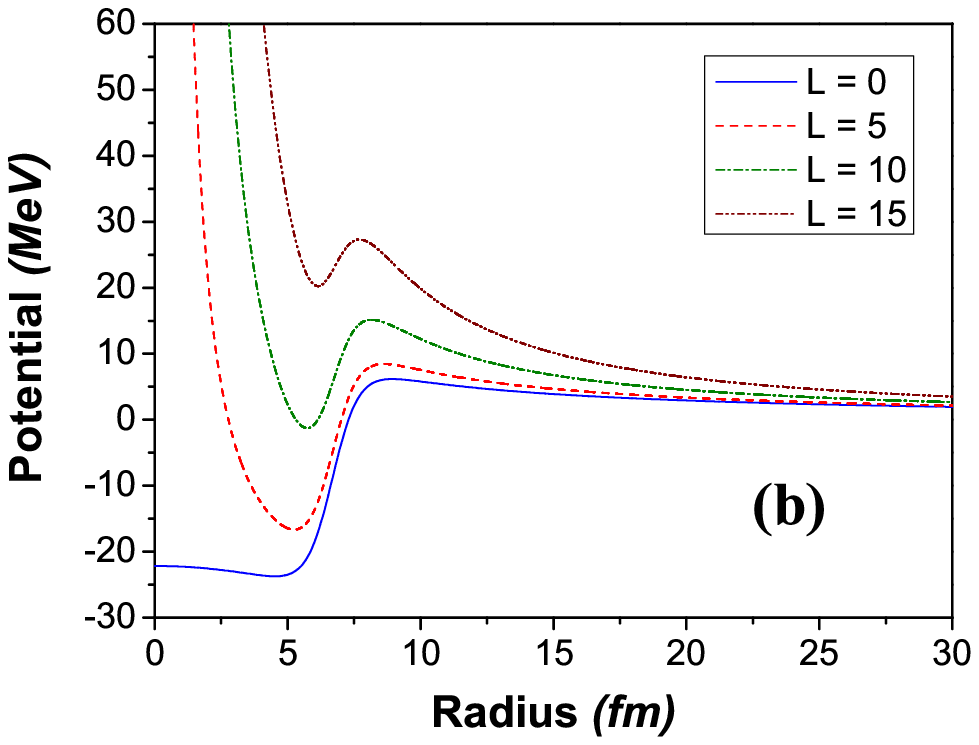}}
\vspace{-5mm}
\caption{\small 
The barrier penetrabilities depend on the energy $E_{\alpha}$ of the $\alpha$-particle (a) and the
corresponding potential barriers (b) at different orbital moments $l$ for the reaction of $\alpha + ^{44}{\rm Ca}$
(parameters of calculations: 10000 intervals at $r_{\rm max}=70$~fm, parametrization from~\cite{Denisov.2005.PHRVA}).
\label{fig.3.2.1}}
\end{figure}

Let us analyze how the penetrability of the barrier depends on the orbital momentum.
The results of such calculations are shown in Fig.~\ref{fig.3.2.1}~(a).
One can see that the line of the penetrability is shifted to the right with increasing orbital momentum $l$. This change is explained by increasing the internal well of the potential before the barrier, resulting in a larger centrifugal component of the potential at larger $l$ (see Fig.~\ref{fig.3.2.1}~(b)). Each subsequent centrifugal component of the $\alpha$-capture cross-section thus gives its own contribution to the full spectrum, starting from some larger energy $E_{\alpha}$. From here, one can conclude that the full $\alpha$-capture cross-section at some sufficiently small energy $E_{\alpha}$ is described by a partial cross-sections at small $l$ only.


\subsection{The fully quantum approach versus the WKB approach in the calculation of the cross-sections
\label{sec.3.3}}

\begin{figure}[htbp]
\centerline{\includegraphics[width=78mm]{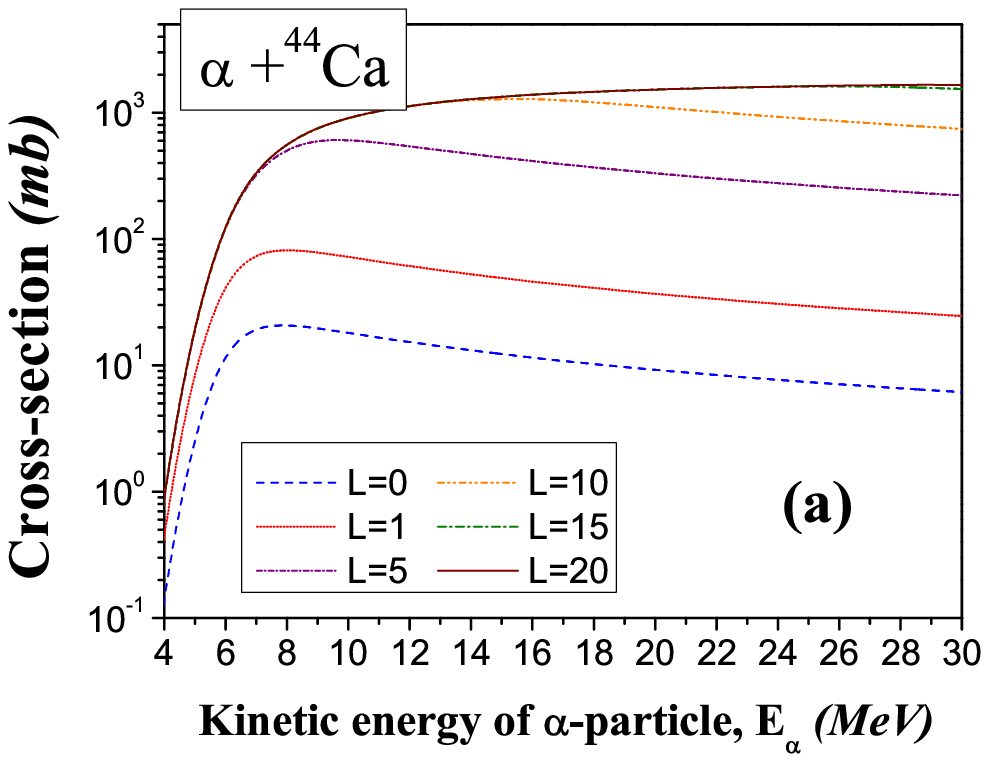}
\hspace{-7mm}\includegraphics[width=78mm]{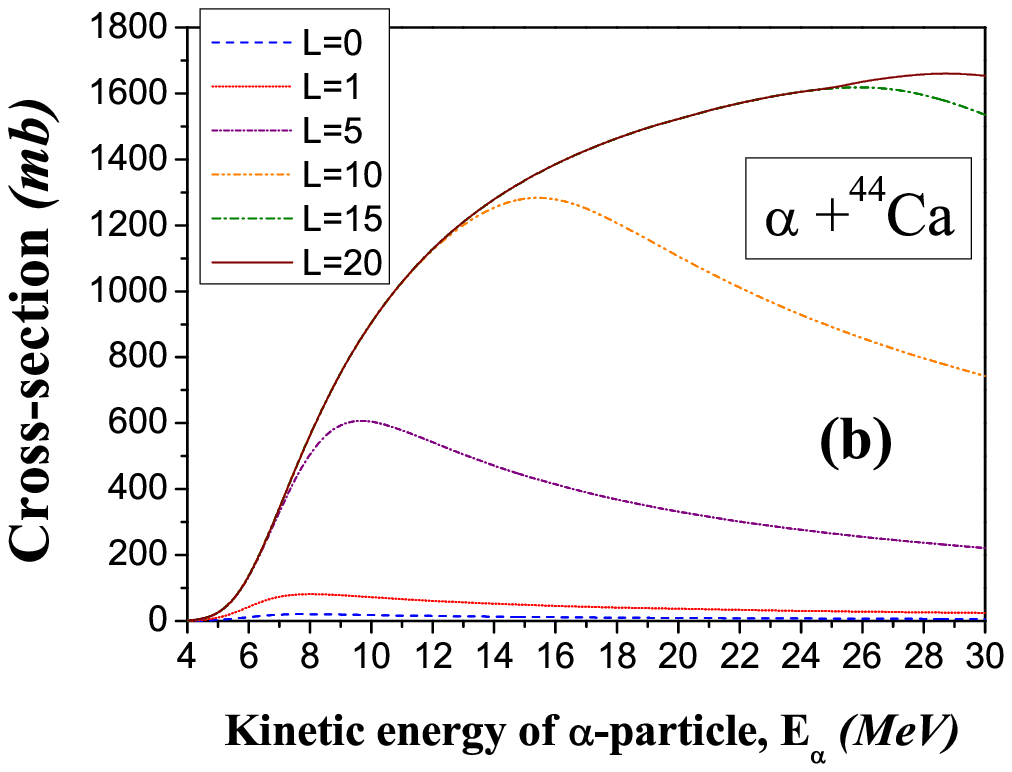}}
\vspace{-5mm}
\caption{\small 
The cross-sections of the capture of the $\alpha$-particle by the $^{44}{\rm Ca}$ nucleus calculated in the MIR approach (parameters of calculations: 10000 intervals at $r_{\rm max}=70$~fm, $r_{\rm capture}$ is chosen at the coordinate of the minimum of the potential well, which changes depending on the used orbital momentum $l$,
parametrization from~\cite{Denisov.2005.PHRVA}).
(a) Spectra in the logarithmic scale: the role of contributions at small $l$ is more visible.
(b) Spectra in the linear scale: the behavior of the summarized spectrum after the inclusion of contributions at large $l$ is more visible.
One can see that each partial cross-section has its maximum. Inclusion of the subsequent contributions (at large $l$) smooths the full spectrum, slowly transforming it into a monotonically increasing curve inside the whole energy range.
\label{fig.3.3.1}}
\end{figure}
Now, we shall consider how the capture cross-section is determined in the fully quantum and WKB approaches.
The results of such calculations in the MIR approach at different values of $L_{\rm max}$ are presented in Fig.~\ref{fig.3.3.1}
(here, $L_{\rm max}$ is the upper limit of the summation in formula~(\ref{eq.2.1.1}), which we choose instead of infinity).
The presentation of the cross-sections in the linear scale clearly shows their slow decrease after the maxima. Additionally, presenting it on a logarithmic scale shows a clear monotonic increase in the cross-sections up to the peaks (especially at small $l$), which clearly demonstrates the behavior of the partial cross-sections at small $l$
on the background of the cross-sections at large $l$.
\begin{figure}[htbp]
\centerline{\includegraphics[width=78mm]{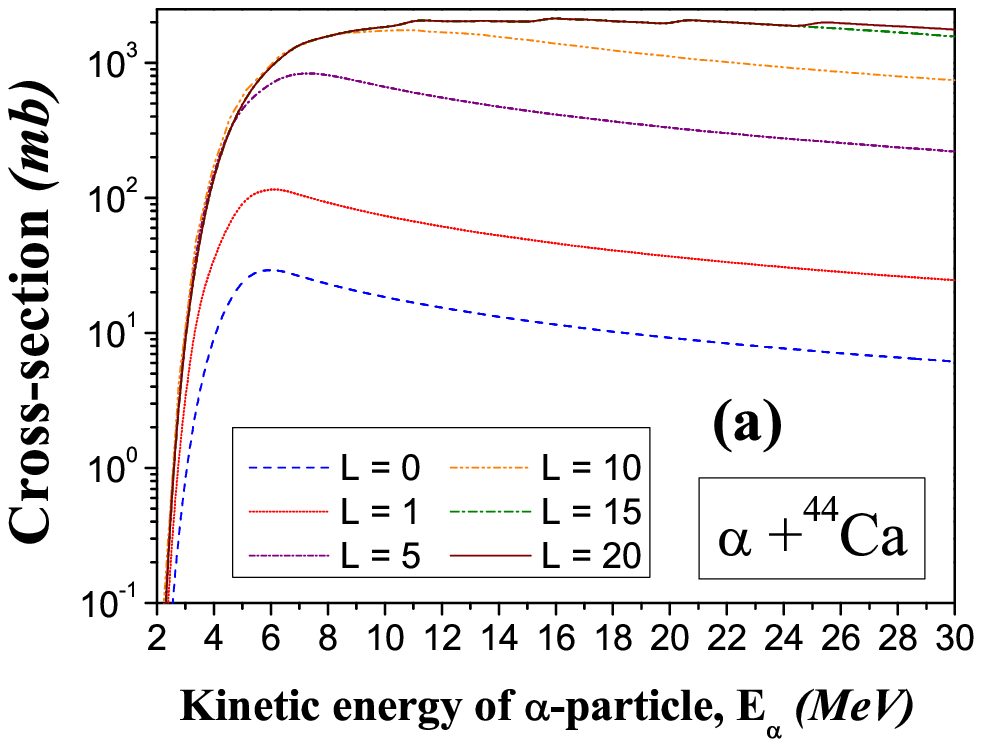}
\hspace{-7mm}\includegraphics[width=78mm]{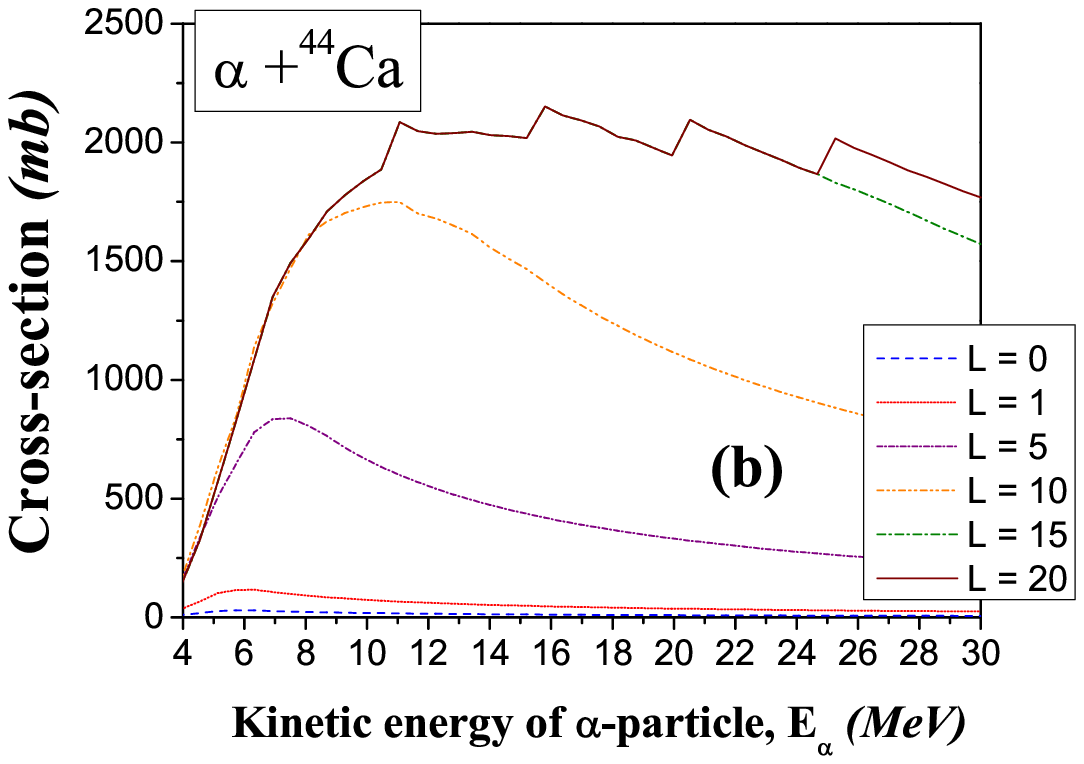}}
\vspace{-5mm}
\caption{\small 
The cross-sections of the capture of the $\alpha$-particle by the $^{44}{\rm Ca}$ nucleus calculated in the WKB approach (parameters of calculations: 10000 intervals at $r_{\rm max}=70$~fm, parametrization from~\cite{Denisov.2005.PHRVA}). In figure~(b), one can see the presence of sharp peaks in the shape of the spectrum that are hidden on a logarithmic scale (see figure~(a)).
Such a peculiarity of the spectrum is natural for the WKB calculations.
\label{fig.3.3.2}}
\end{figure}
The WKB approach gives a different result, shown in Fig.~\ref{fig.3.3.2}. We note the following two aspects (visible in the linear consideration but hidden in the logarithmic):
(1) sharp peaks appear after the inclusion of contributions at large $l$;
(2) the summarized cross-section is sufficiently higher in comparison with the fully quantum calculations presented in Fig.~\ref{fig.3.3.1}.


\subsection{Determination of the probabilities of fusion
\label{sec.3.4}}

\begin{figure}[htbp]
\centerline{\includegraphics[width=155mm]{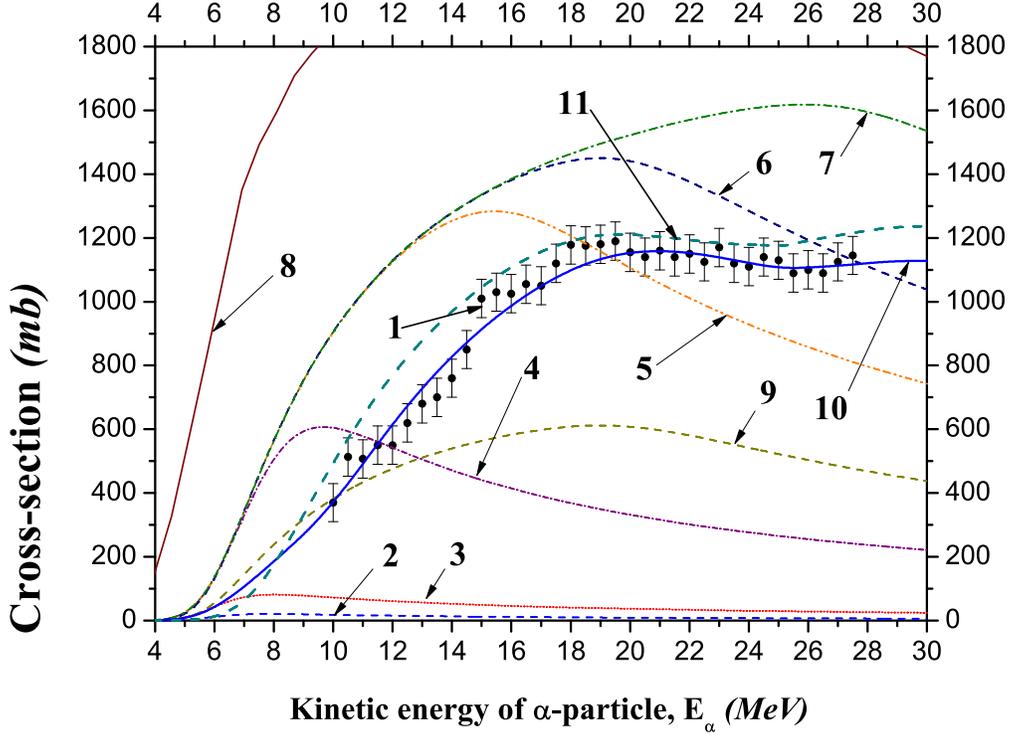}}
\vspace{-9mm}
\caption{\small 
The capture cross-sections of the $\alpha$-particle by the $^{44}{\rm Ca}$ nucleus obtained by the MIR method and WKB approach
(parameters of calculations: 10000 intervals at $r_{\rm max}=70$~fm, parametrization
from~\cite{Denisov.2005.PHRVA}).
Here, the data labeled 1 are the experimental data extracted from \cite{Eberhard.1979.PRL},
dashed blue line 2 is the cross-section at $l_{\rm max}=0$
short dotted red line 3 is the cross-section at $l_{\rm max}=1$,
short dash-dotted purple line 4 is the cross-section at $l_{\rm max}=5$,
dash double-dotted orange line 5 is the cross-section at $l_{\rm max}=10$,
dashed dark blue line 6 is the cross-section at $l_{\rm max}=12$,
dash dotted green line 7 is the cross-section at $l_{\rm max}=15$,
solid brown line 8 is the cross-section at $l_{\rm max}=20$,
dashed dark yellow line 9 is the normalized cross-section at $l_{\rm max}=17$,
solid blue line 10 is the cross-section at $l_{\rm max}=17$, and
dashed dark cyan line 11 is the cross-section at $l_{\rm max}=17$
(cross-sections are defined by (\ref{eq.2.1.1}) where $l_{\rm max}$ is the upper limit in the summation).
The penetrabilities are calculated by the MIR method for lines 2--7 and 9--11, and
by the WKB approach for line 8.
Lines 10-11 are obtained with the included fusion probabilities, and lines 2-9 are obtained without the fusion probabilities.
One can see that line 10, obtained after inclusion of the fusion probabilities, describes
the experimental data with good accuracy (which is not possible without such coefficients but with any arbitrary fitting of the parameter of the $\alpha$-nucleus potential, see Sect.~\ref{sec.3.5}).
For line 11, the fusion probabilities are obtained by formulas~(\ref{eq.3.6.1})--(\ref{eq.3.6.7}).
\label{fig.3.4.1}}
\end{figure}

The cross-sections for $\alpha + ^{44}{\rm Ca}$ obtained by the MIR and WKB approaches and
experimental data~\cite{Eberhard.1979.PRL} are included to Fig.~\ref{fig.3.4.1}.
The MIR calculations with the inclusion of the fusion probabilities found by the minimization method are added to this figure. One can see that the WKB approach gives higher values for the cross-section in comparison with the calculations from the MIR approach at $L_{\rm max} = 10-15$. We conclude the following: (1) \emph{the WKB approach gives reduced estimations for the fusion probabilities in comparison with the MIR approach}, and (2) \emph{the cross-section obtained in the WKB approach has discontinuities at higher energies, while the MIR approach gives the continuous spectrum shape}.

In~\cite{Eberhard.1979.PRL} the \emph{effect of anomalous large-angle scattering (ALAS)} was discussed, which was explained by the \emph{sharp angular momentum cut-off approach at some critical value of $L_{\rm max}$} (see~eqs.(1)--(2) in that paper). This case corresponds to calculations by the MIR approach at different values of $L_{\rm max}$ (where all fusion probabilities are equal to unity at $l \le L_{\rm max}$).
Here, curve 5 for $L_{\rm max} = 10$ in Fig.~\ref{fig.3.4.1} (and curve 6 for $L_{\rm max} = 12$) is clearly better at describing the experimental data than curve 7 for $L_{\rm max} = 15$ (which nearly coincides with subsequent calculations for higher $L_{\rm max}$). However, all these curves are far from the experimental data in comparison with curve 10, which includes the fusion probabilities obtained by the minimization method.
Thus, the sharp angular momentum cutoff approach does not well describe the exiting experimental data for $\alpha + ^{44}{\rm Ca}$, while curve 10 is more successful. This result indicates that the dependence of the fusion probabilities on the angular momentum is more complicated and requires more careful study. The fusion probabilities used in the calculation of curve 10 are presented in Tabl.~\ref{table.app.2.1} in \ref{sec.app.2}. Here, we find that $p_{0}$ \ldots $p_{4}$ are extremely small (in contrast to the sharp angular momentum cutoff approach \cite{Eberhard.1979.PRL}). Additionally, $p_{15}$ and $p_{16}$ are close to unity (at small $p_{13}$), which allows one to describe small oscillations of the spectrum at high energies (at $E=23-28$~MeV). In other words, the presence of the oscillating behavior in the experimental data indicates sufficient high fusion probabilities at maximal angular momenta (in contrast to the sharp angular momentum cutoff approach \cite{Eberhard.1979.PRL}).

\begin{figure}[htbp]
\centerline{\includegraphics[width=92mm]{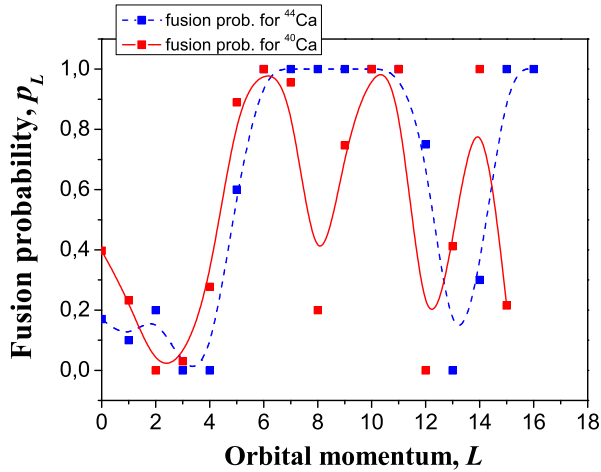}}
\vspace{-5mm}
\caption{\small 
The fusion probabilities for the capture of the $\alpha$-particle by the nuclei $^{40}{\rm Ca}$ (solid red line) and $^{44}{\rm Ca}$ (dashed blue line) obtained by the MIR approach (parameters of calculations: 10000 intervals at $r_{\rm max}=70$~fm, parametrization from~\cite{Denisov.2005.PHRVA}).
\label{fig.3.4.2}}
\end{figure}

Eberhard et al.~\cite{Eberhard.1979.PRL} studied fusion (absorption) of the $\alpha$-particle by the $^{40}{\rm Ca}$ and $^{44}{\rm Ca}$ nuclei and compared the dependencies of the capture cross-sections on the angular momentum $l$. These dependencies were obtained at one chosen energy (in Fig.~2 in that paper, data are presented for $E=25$~MeV). We shall therefore compare the fusion probabilities for these two nuclei as a function of the angular momentum,
and such calculations are presented in Fig.~\ref{fig.3.4.2}.
These data were obtained by analyzing the experimental data inside the whole energy region,
which indicates a more accurate analysis in our approach.
Fig.~\ref{fig.3.4.2} shows that both curves have similar behaviors at low $l$, but they are different at large $l$. The fusion probability for $^{40}{\rm Ca}$ is smaller than that for $^{44}{\rm Ca}$ at large $l$ -- this corresponds to smaller capture cross-sections for $^{40}{\rm Ca}$ than for $^{44}{\rm Ca}$ in the experimental data~\cite{Eberhard.1979.PRL}. Decreasing the capture cross-section for $^{40}{\rm Ca}$ at high energies can thus be explained by a smaller fusion probability at a large $l$ (these two nuclei have different barriers, which give $L_{\rm max}=15$ for $^{40}{\rm Ca}$ and $L_{\rm max}=16$ for $^{44}{\rm Ca}$,
forming an additional contribution to the full spectrum for $^{44}{\rm Ca}$).
The normalization of the calculated spectra on the experimental data (without the inclusion of the fusion probabilities) does a worse job of describing the experimental data (see curve 9 in Fig.~\ref{fig.3.5.1}) than curve 10.


\subsection{Optimal parameters of the nuclear component of the potential
\label{sec.3.5}}

\begin{figure}[htbp]
\centerline{\includegraphics[width=80mm]{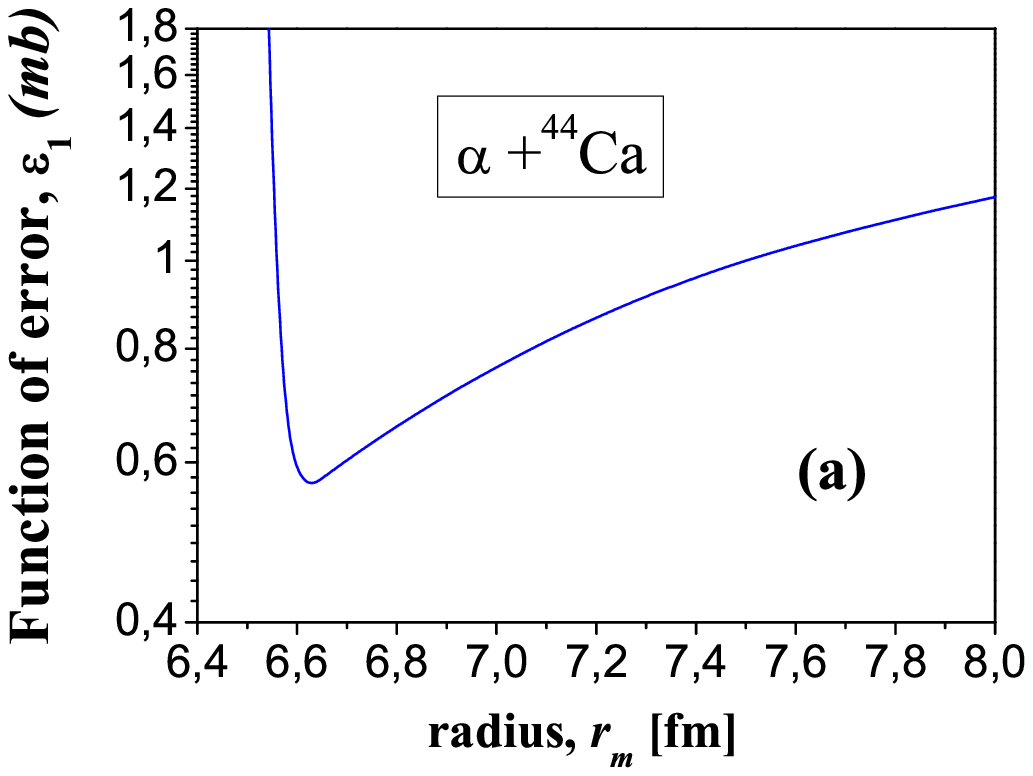}
\hspace{-8mm}\includegraphics[width=78mm]{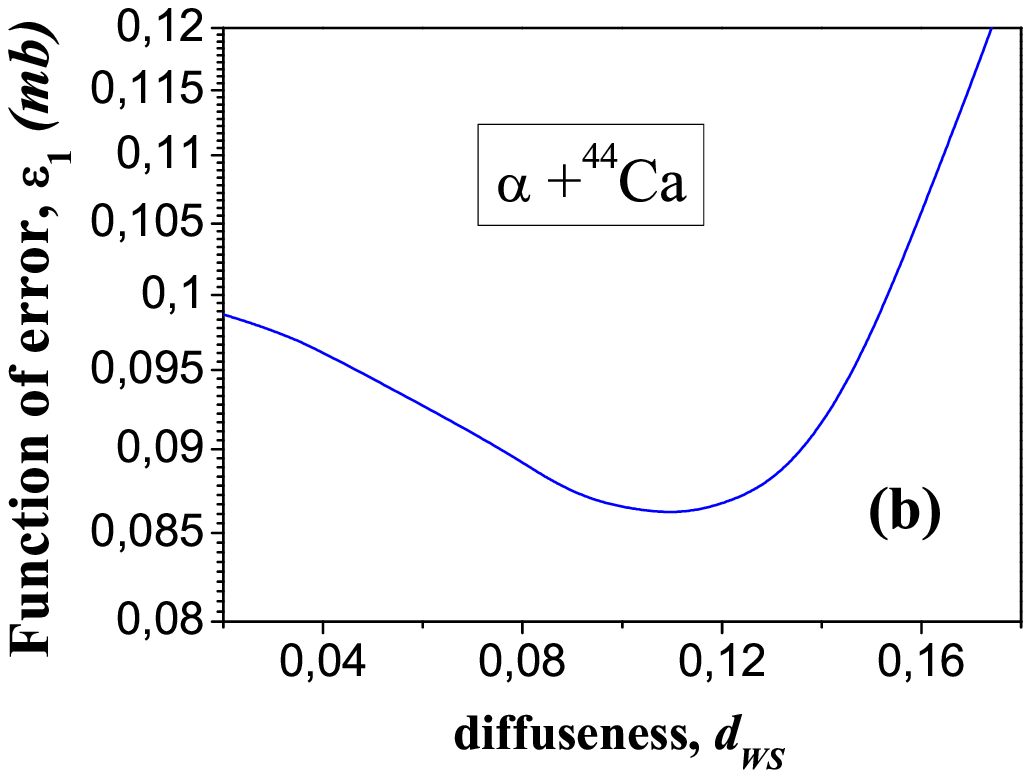}}
\vspace{-5mm}
\caption{\small 
Dependencies of the function of errors $\varepsilon_{1}$ defined by formula~(\ref{eq.2.3.11}) on
the radius $r_{\rm m}$ (a) and
diffuseness $d$ (b) for the capture $\alpha + ^{44}{\rm Ca}$
(parameters of calculations: 10000 intervals at $r_{\rm max}=70$~fm, applied MIR method without inclusion of the fusion probabilities, parametrization from~\cite{Denisov.2005.PHRVA}).
\label{fig.3.5.1}}
\end{figure}

The determination of the potential parameters is a standard task in theory of optimization.
The analysis shows the presence of only one stable minimum in independencies of the function of error on the $r_{m}$ and $d$ parameters (see Fig.~\ref{fig.3.5.1}),
which confirms the applicability of the optimization methods for this task.
%
%
The determination of these parameters by such a method is given in~\ref{sec.app.3} in more details.

In Tabl.~\ref{table.3.5.4} we present calculations by the MIR approach with the included fusion probabilities and the new parameters found by the optimization technique
for the capture of $\alpha + ^{40}{\rm Ca}$ and
$\alpha + ^{44}{\rm Ca}$
and for the WKB approach without the fusion probabilities for parametrization~\cite{Denisov.2005.PHRVA}
(such a situation did exist before our paper).
\begin{table}
\begin{center}
\begin{tabular}{|c|c|c|c|c|c|c|} \hline
 & \multicolumn{3}{|c|}{$\alpha + ^{40}{\rm Ca}$} &
   \multicolumn{3}{|c|}{$\alpha + ^{44}{\rm Ca}$} \\ \hline
  Parametrization & own & \multicolumn{2}{|c|}{Ref.~\cite{Denisov.2005.PHRVA}} &
    own & \multicolumn{2}{|c|}{Ref.~\cite{Denisov.2005.PHRVA}} \\ \hline
  Fusion probabilities & included & \multicolumn{2}{|c|}{excluded} &
                         included & \multicolumn{2}{|c|}{excluded} \\ \hline
  Penetrability app.  & MIRM & WKB & MIRM & MIRM & WKB & MIRM \\ \hline
  $V_{0}$, MeV    & 28.6444 & \multicolumn{2}{|c|}{28.8374} & 36.1   & \multicolumn{2}{|c|}{35.2657} \\
  $r_{\rm m}$, fm &  6.5164 & \multicolumn{2}{|c|}{6.6338} & 7.3    & \multicolumn{2}{|c|}{6.68670} \\
  $d$, fm         &  0.5107 & \multicolumn{2}{|c|}{0.49290} & 0.4349 & \multicolumn{2}{|c|}{0.49290} \\ \hline
  $\varepsilon_{1}$ & 0.02141 & 0.9233 & 0.5154 & 0.0246 & 0.8401 & 0.3881 \\
  $\varepsilon_{2}$ & 0.02716 & 0.5251 & 0.3873 & 0.0344 & 0.4936 & 0.3238 \\
  $\varepsilon_{3}$ & 0.00409 & 0.1030 & 0.0687 & 0.0051 & 0.0957 & 0.0561 \\ \hline
\end{tabular}
\end{center}
\caption{Parameters of the $\alpha$-nucleus potential and functions of error for the capture of $\alpha + ^{40}{\rm Ca}$ and $\alpha + ^{44}{\rm Ca}$.
The new parameter determination results for calculations of the penetrabilities by the MIR method with the included fusion probabilities given in Tabl.~\ref{table.3.5.5} are presented in columns 2 and 5, the MIR calculations of penetrabilities for parametrization~\cite{Denisov.2005.PHRVA} without the included fusion probabilities are given in columns 3 and 6, and calculations of the penetrabilities by the WKB approach at parametrization~\cite{Denisov.2005.PHRVA} without the included fusion probabilities are given in columns 4 and 7.
$\varepsilon_{1}$, $\varepsilon_{2}$ and $\varepsilon_{3}$ are functions of errors defined by the formula~(\ref{eq.2.3.11}).
%
}
\label{table.3.5.4}
\end{table}
One can see that with the same parametrization used in~\cite{Denisov.2005.PHRVA}, the MIR method decreases the error $\varepsilon_{1}$ by $1.79$ times for $\alpha + ^{40}{\rm Ca}$ and in $2.16$ times for $\alpha + ^{44}{\rm Ca}$ in comparison with the WKB calculations.
At the new parametrizations found after the inclusion of the fusion probabilities calculations, the MIR method decreases the error $\varepsilon_{1}$ by $41.72$ times for $\alpha + ^{40}{\rm Ca}$ and by $34.06$ times for $\alpha + ^{44}{\rm Ca}$ in comparison with the WKB calculations at parametrization~\cite{Denisov.2005.PHRVA}.
An analysis of the Hill-Wheeler approach and Wong's formula shows that they are constructed on an extremely strong reduction of the original potential barrier,
and even the WKB approach used a more correct shape of the barrier in the determination of the penetrability
(see~\ref{sec.app.4} for details).
This point is crucial when we extract information about the fusion probabilities from the experimental data.
The updated fusion probabilities for the new parametrization are given in Tabl.~\ref{table.3.5.5}.
\begin{table}
\begin{center}
\begin{tabular}{|c|c|c|c|c|c|c|c|c|c|c|} \hline
  Nucleus & $p_{0}$ & $p_{1}$ & $p_{2}$ & $p_{3}$ & $p_{4}$ & $p_{5}$ & $p_{6}$ & $p_{7}$ & $p_{8}$ & $p_{9}$ \\ \hline
  $^{40}{\rm Ca}$ & 0.05 & 0.01 & 0.01 & 0.29 & 0.89 & 0.94 & 0.60 & 0.43 & 0.83 & 0.60 \\ 
  $^{44}{\rm Ca}$ & 0.04 & 0.03 & 0.01 & 0.01 & 0.43 & 0.31 & 0.62 & 0.31 & 0.71 & 1.00 \\ 
  \hline
\end{tabular}

\vspace{1mm}
\begin{tabular}{|c|c|c|c|c|c|c|c|c|c|} \hline
  Nucleus & $p_{10}$ & $p_{11}$ & $p_{12}$ & $p_{13}$ & $p_{14}$ & $p_{15}$ & $p_{16}$ & $p_{17}$ & $p_{18}$ \\ \hline
  $^{40}{\rm Ca}$ & 1.00 & 1.00 & 0.01 & 0.49 & 0.95 & 0.01 & 0.01 & 0.01 & 0.01 \\ 
  $^{44}{\rm Ca}$ & 1.00 & 1.00 & 1.00 & 0.67 & 0.09 & 0.01 & 0.91 & 1.00 & 0.01 \\ 
  \hline
\end{tabular}
\end{center}
\caption{The fusion probabilities for the capture of $\alpha + ^{40}{\rm Ca}$ and $\alpha + ^{44}{\rm Ca}$ obtained by the MIR approach at parametrization given in Tabl.~\ref{table.3.5.4} (see columns 2 and 5 in that table).
}
\label{table.3.5.5}
\end{table}
These results confirm that the MIR approach (1) is the most accurate at describing the tunneling processes through the original barriers and for calculations of the fusion probabilities in the $\alpha$-capture task,
and (2) allows one to describe the experimental data with the best agreement.

\subsection{The predicted fusion probabilities
\label{sec.3.6}}

From Fig.~\ref{fig.3.4.2} one can find two aspects that suppress fusion during $\alpha$-capture.

1) The first aspect is observed at the initial angular momenta. It essentially suppresses the fusion process after the $\alpha$-particle is transferred the barrier and is put into the spatial internal region of the nucleus. This action could be nuclear attracting forces between the nucleons forming the main contribution of the binding energy of the nucleus and leading to its stability.
However, this action could also be new forces of dissipative nature with the strongest influence on the internuclear motions of nucleons at the smaller angular momenta.
All such forces should have a unified general form for the different nuclei.
From Fig.~\ref{fig.3.4.2} we find the following dependence describing such an aspect:
\begin{equation}
\begin{array}{lcl}
  p_{1} (L) = \displaystyle\frac{c_{1}}{1 + e^{(L - c_{2})/ c_{3}}},
\end{array}
\label{eq.3.6.1}
\end{equation}
\begin{equation}
\begin{array}{lcl}
  c_{1} = 1, & c_{2} = 4.2, & c_{3} = 0.5.
\end{array}
\label{eq.3.6.2}
\end{equation}

2) The second aspect is shown at the final angular momenta. Comparing the curves for the $^{40}{\rm Ca}$ and $^{44}{\rm Ca}$ nuclei in Fig.~\ref{fig.3.4.2}, one can find that this aspect plays a different role for the two nuclei. Therefore, this aspect should be connected with the structure of the nucleus, which can be explained on the basis of the closure of nuclear shells. In particular, this aspect can be the reason for the difference between the spectra for the $^{40}{\rm Ca}$ and $^{44}{\rm Ca}$ nuclei (while the first aspect has the same influence on both nuclei).
From Fig.~\ref{fig.3.4.2}, we suppose the following form of such an aspect:
\begin{equation}
  p_{2} (L) =
  f_{2}(L) \cdot
  \displaystyle\sum\limits_{n=1}
    e^{- \displaystyle\frac{(L - n \cdot \Delta)^{2}}{c_{4n}}}.
\label{eq.3.6.3}
\end{equation}
Here, the function $f_{2}(L)$ suppresses the oscillating dependence with decreasing $L$, and
it can be described as
\begin{equation}
  f_{2}(L) = 1 - e^{- c_{5} \cdot (L - c_{6})}.
\label{eq.3.6.4}
\end{equation}
For the first calculation, we use $c_{4n} = 1$, $c_{5}=0.25$ and $c_{6}=2.5$.
The parameter $\Delta$ should define the influence of the neutrons shells on the calculated fusion probabilities,
which can be described as
\begin{equation}
  \Delta = a \cdot (N - N_{\rm magic}) + b.
\label{eq.3.6.5}
\end{equation}
Here, $N$ is number of neutrons of the studied nucleus, and $N_{\rm magic}$ is the closest magic neutron number, where $N_{\rm magic} \le N$ (i.~e. $N_{\rm magic}=20$ for $^{40}{\rm Ca}$ and $^{44}{\rm Ca}$). We introduce linear interpolation (\ref{eq.3.6.5}) on the basis of the fusion probabilities given in Fig.~\ref{fig.3.4.2}.
Extracting the periods of oscillations from Fig.~\ref{fig.3.4.2}, we obtain:
\begin{equation}
\begin{array}{lcl}
  a = 2.31, & b = 4.05.
\end{array}
\label{eq.3.6.6}
\end{equation}
The complete fusion probability should be written as
\begin{equation}
  p_{\rm full} (L) = 1 - p_{1} (L) - p_{2} (L).
\label{eq.3.6.7}
\end{equation}
However, we recommend such formula for nuclei at close proton shells.

\begin{figure}[htbp]
\centerline{%
\includegraphics[width=78mm]{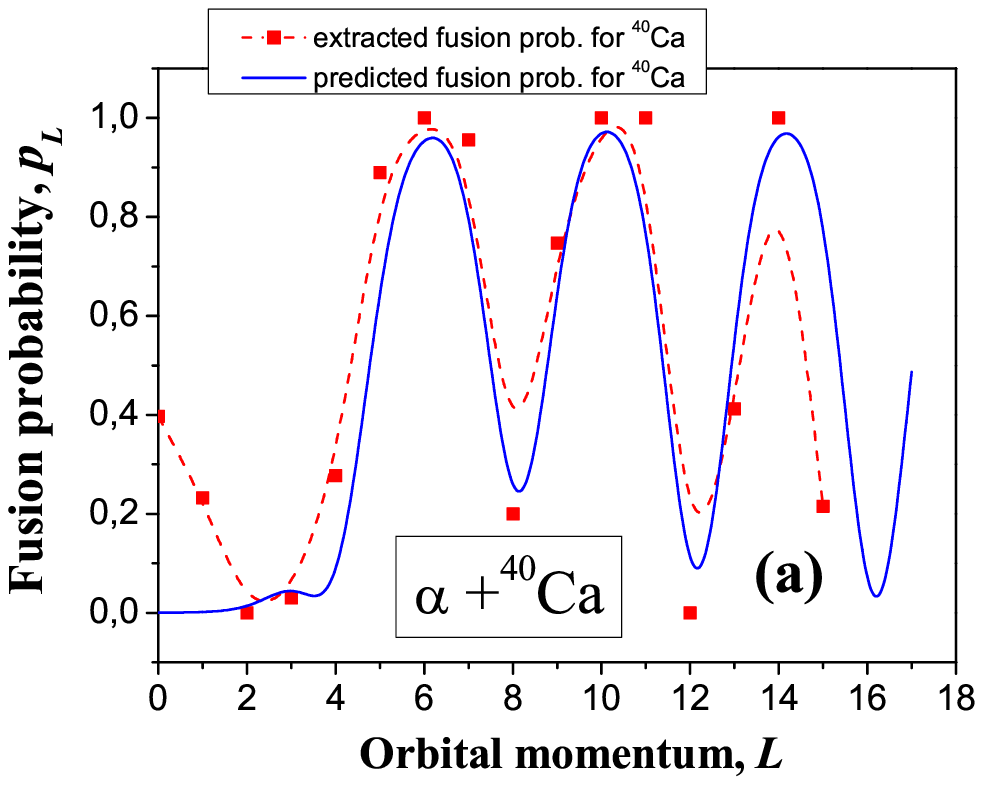}
\hspace{-6mm}\includegraphics[width=78mm]{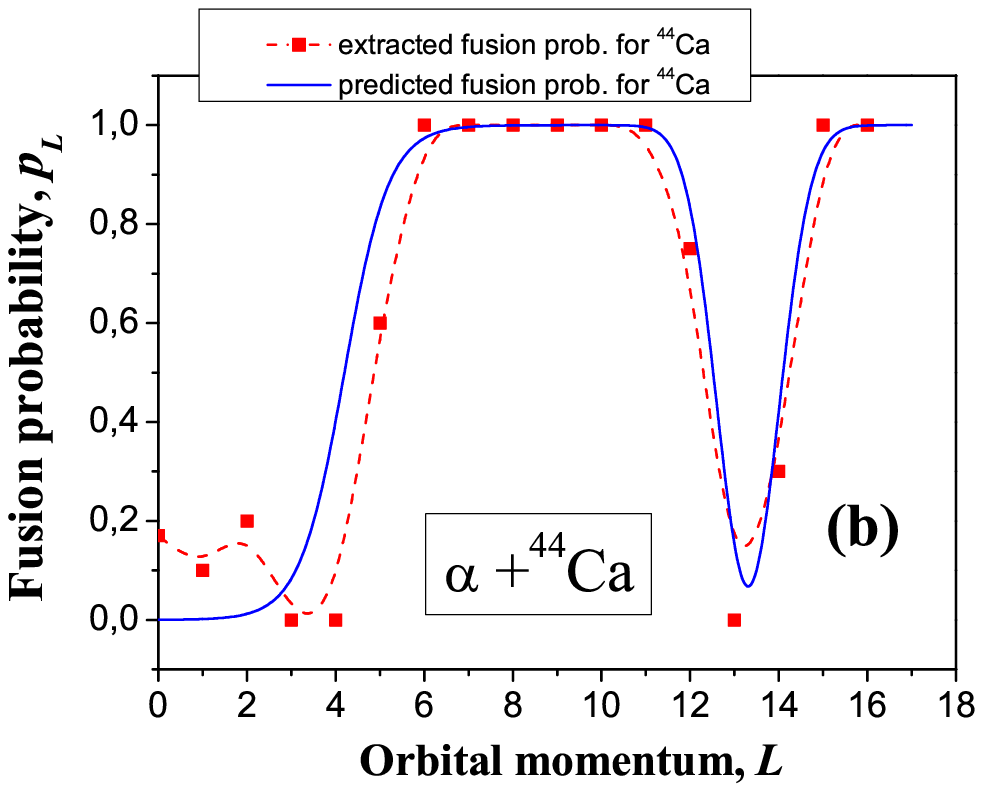}}
\vspace{-5mm}
\caption{\small 
The predicted fusion probabilities for the capture reactions of $\alpha + ^{40}{\rm Ca}$ (a)
and $\alpha + ^{44}{\rm Ca}$ (b).
Here, solid blue lines are calculations for the predicted fusion probabilities given by formulas~(\ref{eq.3.6.1})--(\ref{eq.3.6.7}), and
dashed red lines are the extracted data of the fusion probabilities given in Fig.~\ref{fig.3.4.2}.
\label{fig.3.6.1}}
\end{figure}

The fusion probability calculation results for the two studied nuclei $^{40}{\rm Ca}$ and $^{44}{\rm Ca}$ by these formulas are presented in Fig.~\ref{fig.3.6.1}. One can see that these formulas allow for the sufficient description of the extracted fusion probabilities for both nuclei.
The resulting calculation of the cross-section with the predicted fusion probabilities for
the capture reaction of $\alpha + ^{44}{\rm Ca}$ is added in Fig.~\ref{fig.3.4.1} as the violet dashed line 11.
One can see that the calculated cross-section with the fusion probabilities, when predicted in such a manner, describes the experimental data better
than any other calculation without consideration of fusion during the $\alpha$-capture.

The $^{40}{\rm Ca}$ nucleus is double magic, while the $^{44}{\rm Ca}$ nucleus only has a magic number for protons $Z=20$. Therefore, the binding energy for the $^{40}{\rm Ca}$ nucleus is higher than that for $^{44}{\rm Ca}$. Therefore, the $^{40}{\rm Ca}$ nucleus is more stable and it is more difficult to synthesize a new one by means of the scattering of  the $\alpha$-particle off this nucleus. In particular, the fusion between the $\alpha$-particle and this nucleus during such a scattering reaction should be maximally weakened and, in some cases (for some orbital momenta, etc.), even vanishing. This situation is clearly demonstrated by our results presented in Fig.~\ref{fig.3.4.2}, where one can see smaller fusion probabilities for $^{40}{\rm Ca}$ than for $^{44}{\rm Ca}$ at higher angular momenta. This situation is also confirmed by the experimental cross-sections for $\alpha + ^{40}{\rm Ca}$ and $\alpha + ^{44}{\rm Ca}$.

In 1975, Thibault et al \cite{Thibault.1975.PRC} observed the simultaneous coexistence of two types of shapes in the magic $^{31}{\rm Na}$ nucleus -- spherical and deformed in the ground state. The presence of the deformed shape in the ground state was explained by the strong correlation between $2p$--$2n$ excited energies for the $sd$ and $pf$-shells, which results in an increase of the binding energy and stability reinforcement for the nuclei near the shell at $N=20$. In subsequent studies, it was found that such a coexistence of two shapes is observed for majority of the nuclei near the shells at $N=20$ and $N=28$ \cite{Penionzhkevich.2006.EPAN}.

In this connection, we suppose that \emph{each shape (in the ground state) determines the most stable own state of the nucleus, and it should therefore have a definite relation with the suppression of the fusion processes if one considers the scattering of the $\alpha$-particles on such nuclei}.
In particular, both shapes should appear in the nuclei with a close neutron shell at $N=20$, which one can look for in the experimental data of the $\alpha$-capture reactions. $^{40}{\rm Ca}$ is such a nucleus, for which we see the more essential occurrence of the second suppression aspect of fusion at higher energies of the incident $\alpha$-particles (see Fig.~\ref{fig.3.4.2}).
This process leads to a more essential weakening of the $\alpha$-capture cross-section for such a nucleus at higher energies.

Each shape should be related to its own type of fusion suppression in the $\alpha$-capture reactions for the $^{40}{\rm Ca}$ and $^{44}{\rm Ca}$ nuclei.
In particular, the spherical shape of the nucleus does not involve higher angular momenta, so it should be related to the first aspect (shown at the smallest angular momenta).
The deformed shape can be related to higher angular momenta (that corresponds to higher energies in cross-sections), which is connected with the second aspect of reducing the fusion.
We observe the coexistence of two such peculiarities in our results in Fig.~\ref{fig.3.4.2}, which are correlated with the experimental data for both nuclei.
By such logic, a deeper study of the coexistence of the spherical and deformed shapes of the nuclei in the ground state can be recommended via reactions of the $\alpha$-capture by the interesting nuclei (both theoretically and experimentally).

\begin{figure}[htbp]
\centerline{%
\includegraphics[width=78mm]{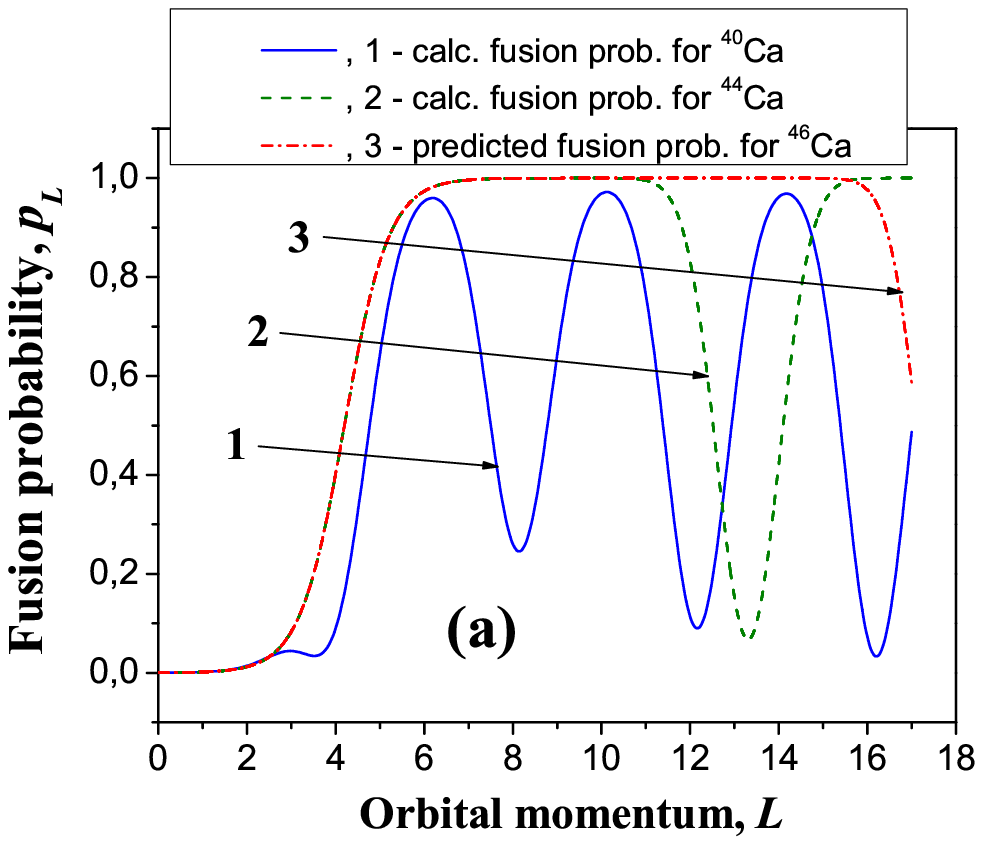}
\hspace{-8mm}\includegraphics[width=85mm]{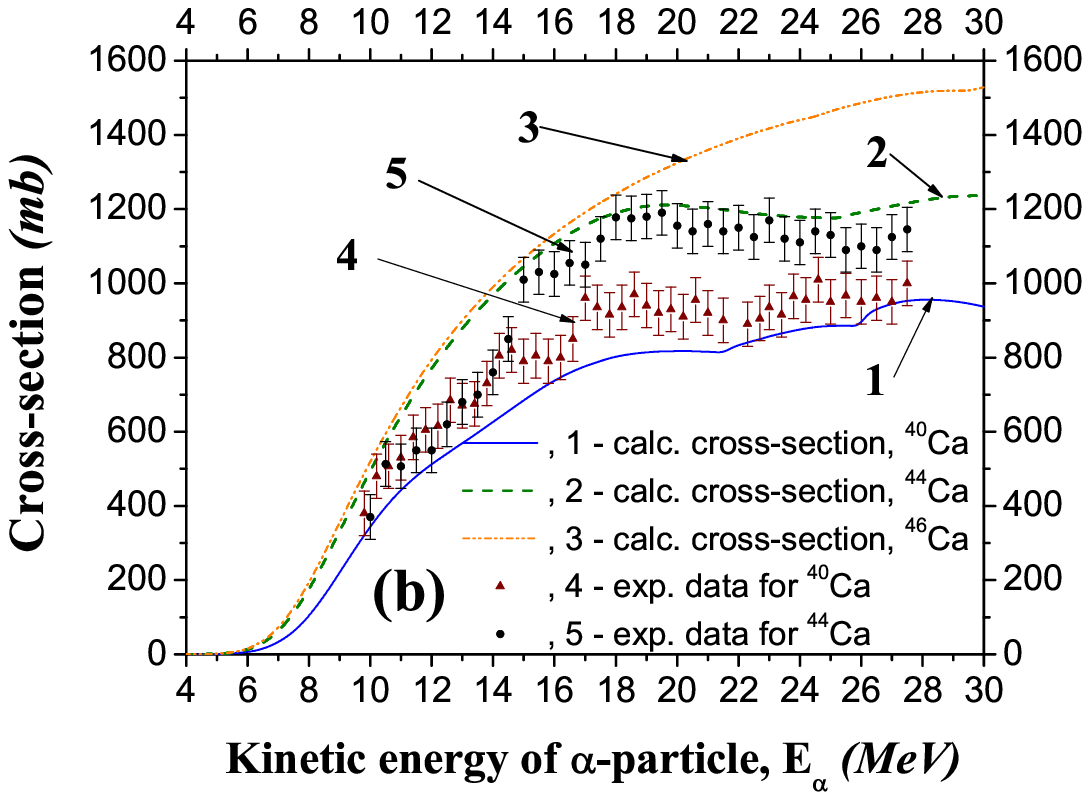}}
\vspace{-5mm}
\caption{\small 
The fusion probabilities calculated by formulas (\ref{eq.3.6.1})--(\ref{eq.3.6.7}) (a) and cross-sections (b) for capture of the $\alpha$-particle by the $^{40}{\rm Ca}$, $^{44}{\rm Ca}$ and $^{46}{\rm Ca}$ nuclei obtained by the MIR method
(parameters of calculations: 10000 intervals at $r_{\rm max}=70$~fm).
\label{fig.6.3.2}}
\end{figure}

Confirmation of the coexistence of the two shapes for the nuclei near the shells at $N=20$ and $N=28$ has lead to the reconsideration of the main positions of the nuclear shell model and to the discovery of new magic numbers \cite{Penionzhkevich.2006.EPAN}.
A review of this topic \cite{Penionzhkevich.2006.EPAN} gives interesting indications about the new neutron magic numbers at $N=16$ and $N=26$ and the properties of such nuclei (while the standard theory gives us only seven experimentally known neutron numbers at 2, 8, 20, 28, 50, 82, 126). In this regard, it could be interesting for experimentalists to propose to investigate the fusion process at the capture of the $\alpha$-particle by the $^{46}{\rm Ca}$ nucleus. Such information could provide new insight into our understanding of physics of nuclei with such a neutron magic shell.
We support such research by our predictions for this nucleus given in Fig.~\ref{fig.6.3.2}.

\section{Conclusions
\label{sec.conclusions}}

In this paper, the multiple internal reflections method is generalized in the description of the capture of $\alpha$-particles by nuclei.
On this basis, the $\alpha$-captures by the $^{40}{\rm Ca}$ and $^{44}{\rm Ca}$ nuclei are analyzed.
Note the following:

\begin{enumerate}
\item
Comparing the capture cross-sections calculated by the MIR method with the experimental data \cite{Eberhard.1979.PRL}, we find
(1) new fusion probabilities of the $\alpha$-particles with the $^{40}{\rm Ca}$ and $^{44}{\rm Ca}$ nuclei (see Tabl.~\ref{table.3.5.5}), and
(2) new parameters of the $\alpha$--nucleus potential (see Tabl.~\ref{table.3.5.4}).
Considering the fusion processes and including them in the calculations essentially allows us to increase agreement with experimental data (see Tabl.~\ref{table.3.5.4}).
In particular, the MIR method found new parametrizations with the included fusion probabilities and
decreased the error $\varepsilon_{1}$ by $41.72$ times for $\alpha + ^{40}{\rm Ca}$ and by $34.06$ times for $\alpha + ^{44}{\rm Ca}$ in comparison with the WKB calculations for parametrization~\cite{Denisov.2005.PHRVA} without the included fusion probabilities.

\item
The found fusion probabilities indicate that the sharp angular momentum cutoff approach proposed by Glas and Mosel in~\cite{Glas.1975.NPA, Glas.1974.PRC} is a rough approximation (see curve 6 in Fig.~\ref{fig.3.5.1}), which excludes the possibility of studying fusion in the $\alpha$-capture reactions (see \ref{sec.app.2} for explanations).
A more realistic picture is given in Table~\ref{table.app.2.1} (corresponding to curve 10 in Fig.~\ref{fig.3.5.1}).
In particular, the fusion probabilities at the first angular momenta are close to zero,
contradicting the idea of a sharp angular momentum cutoff (see Tabl.~\ref{table.3.5.5}).

\item
It is shown that (1) the WKB approximation gives reduced probabilities of fusion in comparison with the MIR approach, and (2) the capture cross-section obtained in the WKB approach has discontinuities at higher energies (the MIR approach provides a continuous spectrum shape).

\item
It is shown that Wong's formula and the Hill-Wheeler approach for barrier penetrability determination
(1) use essentially a reduced inverse parabolic approximated shape of the original barrier,
(2) do not use the shape of the realistic barrier (with the exception of the localized region near the barrier maximum point), and
(3) do not correspond to standard penetrability in quantum mechanics.

\item
A new formula of the fusion probabilities is proposed.
Explanations of the differences between the cross-sections for $\alpha + ^{40}{\rm Ca}$ and $\alpha + ^{44}{\rm Ca}$ and the previously discovered coexistence of the spherical and deformed shapes in the ground state for nuclei near the neutron magic shell $N=20$ are given by our formula for fusion probabilities.
To obtain deeper insight into the physics of nuclei with the new magic number $N=26$, a cross-section of $\alpha + ^{46}{\rm Ca}$ is predicted for further experimental confirmation.

\item
We compared the effectiveness of the MIR approach in determining of the penetrability and reflection coefficients
with an approach based on the determination of the wave functions obtained via the direct integration of the radial Schr\"{o}dinger equation with high accuracy using the Numerov technique.
For analysis, we chose the reaction $\alpha + ^{44}{\rm Ca}$ for case $l=0$.
We conclude the following (see~\ref{sec.app.5}):
\begin{enumerate}
\item
To estimate the accuracy of the determination of the penetrability and reflection coefficients, we use the standard test of quantum mechanics, calculating parameter $\varepsilon_{1} = ||T + R| - 1|$ by each method (here, $T$ and $R$ are the penetrability and reflection coefficients).
The Numerov approach determines these coefficients with accuracy $\varepsilon_{1} = 10^{-5}$, while the accuracy of the MIR approach is restricted by computer capability (see Tables~\ref{table.app.5.1} and~\ref{table.app.5.2}, where we provide the MIR calculations at $\varepsilon_{1} \le 10^{-16}$ for any chosen $r_{\rm max}$ and $N$).

\item
Analyzing the stability of determination of the penetrability and reflection coefficients after varying the parameters $r_{\rm max}$ and $N$, we conclude that the MIR approach is essentially more accurate than the Numerov approach.
The accuracy of determination of these coefficients by the Numerov approach is restricted by limit $10^{-3}$: we have only 3 stable digits for these coefficients at variation of $r_{\rm max}$ and number of intervals $N$.
Simultaneously, the accuracy of the determination of these coefficients by the MIR approach is infinitely increased by increasing the number of intervals $N$:
we reach the first 8 stable digits of these coefficients when $N \to 50000$
(see Tables~\ref{table.app.5.1} and~\ref{table.app.5.2}).
\end{enumerate}

In summary, we conclude that the MIR approach is more accurate and stable. Simultaneously, it requires a similar time for calculations as the Numerov approach.

This result can be explained by the following.
The MIR approach gives exact analytical solutions for the wave function (for the chosen boundary conditions) for the potential with any finite number of rectangular steps. The accuracy of the description of the original potential by the rectangular steps potential is determined only by the number of steps, where the limit of increasing the number of such steps is restricted only by computer capability (i.e., in principle, the methodological accuracy of such an approach can be increased up to infinity by the presented formalism in this paper).

At the same time, the accuracy of the Numerov approach has some limits that cannot be overcame.
The accuracy of such an approach is restricted by the following:
(1) The Numerov technique gives values for the wave function in the next radial coordinate on the basis of the values of the wave function in the previous two radial coordinates with some error, i.e., it is approximated.
Such accuracy is only valid inside a sufficiently small radial region.
(2) Error in the determination of the wave function inside the whole radial region of action of nuclear forces on the basis of the Numerov technique is increased (with increasing number of intervals $N$).
(3) The amplitudes of the Coulomb functions at the near asymptotic region are calculated on the basis of asymptotic series that are divergent
(usually, such functions are used at the best convergent limit).

\item
We analyzed the role of the nuclear deformations in the determinations of the fusion probabilities in the $\alpha$-capture task for $\alpha + ^{44}{\rm Ca}$.
We find the following (see~\ref{sec.app.6}):
\begin{enumerate}
\item
Analysis of changes in the penetrability of the potential barriers after the inclusion of the deformation of the nucleus for case $l=0$ provides the clearest understanding of such a question.
The largest influence of the nuclear deformation exists for angles of deformation $\theta$ of $0^{\circ}$ and $90^{\circ}$ and for the lowest energy of the $\alpha$-particle, while at $\theta = 55^{\circ}$, such an influence is missing (see Fig.~\ref{fig.app.6.1}).

\item
The MIR approach allows one to obtain the fusion probabilities for the deformed nuclei and to describe the experimental data of the $\alpha$-capture cross-sections with high accuracy similar to that for the spherical nuclei (see Tabl.~\ref{table.app.6.1} and lines 2 and 3 in Fig.~\ref{fig.app.6.2}).
We observe the presence of small oscillations in the calculated spectrum after the inclusion of the nuclear deformation.
\end{enumerate}

Another important aspect is the determination of the boundary where the transmitted flux is defined. It turns out that small displacements of this boundary essentially change the resulting penetrability and reflection coefficients.
For example, for $\alpha + ^{44}{\rm Ca}$ at $l=0$ and $E_{\alpha}=2$~MeV the
role of such a displacement can be estimated by the ratio
$f_{1} = T (r_{\rm capture}=7.42~{\rm fm}) / T (r_{\rm capture}=4.5~{\rm fm}) =
4.09869 \cdot 10^{-5} / 2.07705 \cdot 10^{-5} = 1.9733$
(see Fig.~1).
Simultaneously, the role of nuclear deformation in this task can be estimated by the ratio of the penetrabilities for spherical and deformed nuclei for this reaction.
We have $(T_{\rm def} - T_{\rm spher}) / T_{\rm spher} = 0.58015$ for $E_{\alpha}=2.1$~MeV and $l=0$ (see Fig.~F.12~(b) at close energies).
From here, we obtain the ratio
$f_{2} = T_{\rm def} / T_{\rm spher} =
(T_{\rm def} - T_{\rm spher}) / T_{\rm spher} + 1 = 0.58015 + 1 = 1.58015$.
One can see that this ratio $f_{2}$ is less than the ratio $f_{1}$.
From here, one can conclude that in fully quantum calculations, the role of the position of the boundary $r_{\rm capture}$ in determining the resulting penetrability (and cross-sections) is more important than the nuclear deformations.
To study such an aspect correctly in the $\alpha$-capture task (and to work with it), some formalism should be constructed. We use the MIR approach for this aim.
This is one of reasons to use the MIR technique.

\end{enumerate}
We summarize that the fusion processes are an essential ingredient in nuclear capture, and we propose the use of the MIR method as a fully quantum highly precision tool for the study of such processes in nuclear tasks.


\section*{Acknowledgements
\label{sec.acknowledgements}}

The authors are grateful to Prof.~A.~G.~Magner for useful discussions concerning calculations of the penetrabilities through inverse parabolic nuclear barriers.
S.~P.~Maydanyuk thanks the Institute of Modern Physics of the Chinese Academy of Sciences for its warm hospitality and support.
This work was supported by the Major State Basic Research Development Program in China (No. 2015CB856903), the National Natural Science Foundation of
China (Grant Nos. 11035006 and 11175215), and
the Chinese Academy of Sciences fellowships for researchers from developing countries (No. 2014FFJA0003).



\appendix

\section{Direct method
\label{sec.app.1}}

We shall soon add a solution for the amplitudes of the wave function obtained by the standard technique of quantum mechanics that could be obtained if used only in the condition of continuity of the wave function and its derivative at each boundary, but on the whole region of the studied potential (key relations were given in~\cite{Maydanyuk.2003.PhD-thesis}; such an approach was further applied for the study of $\alpha$-decay, proton decay~\cite{Maydanyuk.2011.JMP}, and tunneling in quantum cosmology tasks \cite{Maydanyuk.2010.IJMPD,Maydanyuk.2011.EPJP}). First, we find the functions $f_{N-1}$ and $g_{N-1}$
(from the last boundary at $r_{N-1}$):
\begin{equation}
\begin{array}{cc}
  f_{N-1} = \displaystyle\frac{k_{N-1}+k_{N}}{k_{N-1}-k_{N}} \,e^{2ik_{N-1}r_{N-1}}, &
  g_{N-1} = \displaystyle\frac{2k_{N}}{k_{N}-k_{N-1}} \,e^{i(k_{N}+k_{N-1})r_{N-1}}.
\end{array}
\label{eq.app.1.1}
\end{equation}
Then, using the following recurrent relations:
\begin{equation}
\begin{array}{ccl}
  f_{j-1} & = & \displaystyle\frac
              {(k_{j-1}-k_{j})\, e^{2ik_{j}r_{j-1}} + f_{j}\, (k_{j-1}+k_{j}) }
              {(k_{j-1}+k_{j})\, e^{2ik_{j}r_{j-1}} + f_{j}\, (k_{j-1}-k_{j}) }
              \cdot e^{2ik_{j-1}r_{j-1}},
\end{array}
\label{eq.app.1.2}
\end{equation}
we calculate the subsequent functions $f_{N-2}$, $f_{N-3}$, $f_{N-4}$ \ldots $f_{1}$,
and by such a formula:
\begin{equation}
\begin{array}{ccl}
  g_{j-1} & = & g_{j} \cdot \displaystyle\frac{2k_{j}\, e^{i(k_{j-1}+k_{j})r_{j-1}}}
              {(k_{j-1}+k_{j})\, e^{2ik_{j}r_{j-1}} + f_{j}\, (k_{j-1}-k_{j})}
\end{array}
\label{eq.app.1.3}
\end{equation}
the functions $g_{N-2}$, $g_{N-3}$, $g_{N-4}$ \ldots $g_{1}$.
From $f_{1}$ and $g_{1}$, we find amplitudes $\alpha_{1}$ and $\beta_{1}$ and the amplitude of
transmission $A_{T}$:
\begin{equation}
\begin{array}{cc}
  \alpha_{1} = 0, &
  A_{T} = \beta_{1} = -\displaystyle\frac{g_{1}} {f_{1}}.
\end{array}
\label{eq.app.1.4}
\end{equation}
Now, using the recurrent relations:
\begin{equation}
\begin{array}{ccl}
  \beta_{j+1} & = & \displaystyle\frac
          {\beta_{j}\, e^{ik_{j}r_{j}} + \alpha_{j}\, e^{-ik_{j}r_{j}} - g_{j+1}\, e^{-ik_{j+1}r_{j}}}
          {e^{ik_{j+1}r_{j}} + f_{j+1}\, e^{-ik_{j+1}r_{j}}}
\end{array}
\label{eq.app.1.5}
\end{equation}
and such a formula:
\begin{equation}
  \alpha_{j} = \beta_{j} \cdot f_{j} + g_{j},
\label{eq.app.1.6}
\end{equation}
we consistently calculate the amplitudes $\alpha_{2}$, $\beta_{2}$, $\alpha_{3}$, $\beta_{3}$ \ldots $\alpha_{N-1}$, $\beta_{N-1}$.
Finally, we find the amplitude of reflection $A_{R}$:
\begin{equation}
  A_{R} = \beta_{N-1}\,e^{i(k_{N}+k_{N1})r_{N-1}} + \alpha_{N-1}\,e^{i(k_{N}-k_{N-1})r_{N-1}} - e^{2 ik_{N}r_{N-1}}.
\label{eq.app.1.7}
\end{equation}
As a test, we use condition:
\begin{equation}
  \displaystyle\frac{k_{1}}{k_{N}}\; |A_{T}|^{2} + |A_{R}|^{2} = 1.
\label{eq.app.1.8}
\end{equation}
Studying the problem of proton decay, we used a technique to check the amplitudes obtained previously by the MIR approach and obtained coincidence up to the first 15 digits for all considered amplitudes. \emph{The results for the dependence of the penetrability of the position of point $r_{\rm capture}$ and the obtained cross-sections are thus independent of the fully quantum method used}.


\section{Calculations of the fusion probabilities
\label{sec.app.2}}

In Tabl.~\ref{table.app.2.1}, we present the results of the fusion probabilities calculations for the capture reaction $\alpha + ^{44}{\rm Ca}$, with in dependence on the different maximal values of the orbital momentum $L_{\rm max}$ (and the corresponding functions of errors).
Here, the penetrabilities of the barrier are calculated by the MIR method,
and the parametrization of the $\alpha$-nucleus potential is used from~\cite{Denisov.2005.PHRVA}.
\begin{table}
\hspace{-2mm}
\begin{tabular}{|c|c|c|c|c|c|c|c|c|c|c|c|c|c|c|c|c|} \hline
 & \multicolumn{12}{|c|}{$L_{\rm max}$} \\
 \cline{2-13}
  & 
            5 & 6 & 7 & 8 & 9 & 10 & 11 & 12 & 13 & 14 & 15 & 16 \\ \hline
  $p_{0}$ & 
            1.0 & 1.0 & 1.0 & 1.0 & \hspace{-1mm}0.01\hspace{-1.3mm} &
            \hspace{-1mm}0.01\hspace{-1.3mm} & 0 & 0 & 0.01 & 0.01 & 0.005 & 0.156 \\ 
  $p_{1}$ & 
            1.0 & 1.0 & 1.0 & 1.0 & \hspace{-1mm}0.16\hspace{-1.3mm} &
            0 & 0 & 0 & 0.01 & 0 & 0.001 & 0.090 \\ 
  $p_{2}$ & 
            1.0 & 1.0 & 1.0 & 1.0 & 1.0 &
            0 & 0 & 0 & 0.02 & 0 & 0.001 & 0.175 \\ 
  $p_{3}$ & 
            1.0 & 1.0 & 1.0 & 1.0 & 1.0 &
            0 & 0 & 0 & 0.02 & 0.02 & 0.222 & 0.204 \\ 
  $p_{4}$ & 
            1.0 & 1.0 & 1.0 & 1.0 & 1.0 &
            \hspace{-1mm}0.58\hspace{-1.3mm} & 0 & 0 & 0.26 & 0.19 & 0.415 & 0.190 \\ 
  $p_{5}$ & 
            1.0 & 1.0 & 1.0 & 1.0 & 1.0 & 1.0 & \hspace{-1mm}0.73\hspace{-1.3mm} &
            \hspace{-1mm}0.45\hspace{-1.3mm} & 0.86 & 0.81 & 0.492 & 0.458 \\ 
  $p_{6}$ & 
            - & 1.0 & 1.0 & 1.0 & 1.0 & 1.0 & 1.0 & 1.0 & 0.22 & 0.36 & 0.332 & 0.455 \\ 
  $p_{7}$ & 
            - & - & 1.0 & 1.0 & 1.0 & 1.0 & 1.0 & 1.0 & 1.0 & 1.0 & 1.0 & 1.0 \\ 
  $p_{8}$ & 
            - & - & - & 1.0 & 1.0 & 1.0 & 1.0 & 1.0 & 1.0 & 1.0 & 1.0 & 1.0 \\ 
  $p_{9}$ & 
            - & - & - & - & 1.0 & 1.0 & 1.0 & 1.0 & 1.0 & 1.0 & 1.0 & 1.0 \\ 
  $p_{10}$ & 
            - & - & - & - & - & 1.0 & 1.0 & 1.0 & 1.0 & 1.0 & 1.0 & 1.0 \\ 
  $p_{11}$ & 
            - & - & - & - & - & - & 1.0 & 1.0 & 1.0 & 1.0 & 1.0 & 1.0 \\ 
  $p_{12}$ & 
            - & - & - & - & - & - & - & 1.0 & 0.75 & 1.0 & 1.0 & 1.0 \\ 
  $p_{13}$ & 
            - & - & - & - & - & - & - & - & 1.0 & 0.06 & 0.32 & 0.277 \\ 
  $p_{14}$ & 
            - & - & - & - & - & - & - & - & - & 0.86 & 0 & 0. \\ 
  $p_{15}$ & 
            - & - & - & - & - & - & - & - & - & - & 1.0 & 1.0 \\ 
  $p_{16}$ & 
            - & - & - & - & - & - & - & - & - & - & - & 1.0 \\ \hline
  $\varepsilon_{1}$ & 
          \hspace{-1.5mm}0.540\hspace{-1.9mm} &
          \hspace{-1.5mm}0.471\hspace{-1.9mm} &
          \hspace{-1.5mm}0.411\hspace{-1.9mm} & 
          \hspace{-1.5mm}0.347\hspace{-1.9mm} &
          \hspace{-1.5mm}0.283\hspace{-1.8mm} &
          \hspace{-1.5mm}0.209\hspace{-1.8mm} & 
          \hspace{-1.5mm}0.129\hspace{-1.8mm} &
          \hspace{-1.5mm}0.0569\hspace{-1.8mm} &
          \hspace{-1.5mm}0.0364\hspace{-1.8mm} &
          \hspace{-1.5mm}0.0314\hspace{-1.8mm} & 
          \hspace{-1.5mm}0.0301\hspace{-1.8mm} & 
          \hspace{-1.5mm}0.0293\hspace{-1.8mm} \\ \hline
  $\varepsilon_{2}$ & 
          \hspace{-1.5mm}1.962\hspace{-1.9mm} &
          \hspace{-1.5mm}1.254\hspace{-1.9mm} &
          \hspace{-1.5mm}0.825\hspace{-1.9mm} & 
          \hspace{-1.5mm}0.545\hspace{-1.9mm} &
          \hspace{-1.5mm}0.363\hspace{-1.8mm} &
          \hspace{-1.5mm}0.298\hspace{-1.8mm} & 
          \hspace{-1.5mm}0.172\hspace{-1.8mm} &
          \hspace{-1.5mm}0.0735\hspace{-1.8mm} &
          \hspace{-1.5mm}0.0472\hspace{-1.8mm} &
          \hspace{-1.5mm}0.0420\hspace{-1.8mm} & 
          \hspace{-1.5mm}0.0405\hspace{-1.8mm} & 
          \hspace{-1.5mm}0.0397\hspace{-1.8mm} \\ \hline
  $\varepsilon_{3}$ & 
          \hspace{-1.5mm}0.171\hspace{-1.9mm} &
          \hspace{-1.5mm}0.125\hspace{-1.9mm} &
          \hspace{-1.5mm}0.095\hspace{-1.9mm} & 
          \hspace{-1.5mm}0.071\hspace{-1.9mm} &
          \hspace{-1.5mm}0.054\hspace{-1.8mm} &
          \hspace{-1.5mm}0.041\hspace{-1.8mm} & 
          \hspace{-1.5mm}0.024\hspace{-1.8mm} &
          \hspace{-1.5mm}0.0107\hspace{-1.8mm} &
          \hspace{-1.5mm}0.0070\hspace{-1.8mm} &
          \hspace{-1.5mm}0.0062\hspace{-1.8mm} & 
          \hspace{-1.5mm}0.0060\hspace{-1.8mm} & 
          \hspace{-1.5mm}0.0059\hspace{-1.8mm} \\ \hline
  $E_{\rm min}$ & 
  \hspace{-3mm}-16.6\hspace{-4mm} & \hspace{-3mm}-14.1\hspace{-3mm} & \hspace{-3mm}-11.3\hspace{-4mm} &
  \hspace{-3mm}-8.26\hspace{-4mm} & \hspace{-3mm}-4.90\hspace{-4mm} & \hspace{-3mm}-1.29\hspace{-4mm} &
  \hspace{-3mm}2.56\hspace{-4mm} & \hspace{-3mm}6.50\hspace{-3mm} & 10.9 & 15.5 & 20.24 & 25.15 \\ \hline
  $E_{\rm max}$ & 
  \hspace{-3mm}8.45\hspace{-3mm} & \hspace{-3mm}9.41\hspace{-3mm} & \hspace{-3mm}10.5\hspace{-3mm} &
  \hspace{-3mm}11.8\hspace{-3mm} & \hspace{-3mm}13.4\hspace{-3mm} & \hspace{-3mm}15.5\hspace{-3mm} &
  \hspace{-3mm}17.1\hspace{-3mm} & \hspace{-3mm}19.3\hspace{-3mm} & \hspace{-3mm}21.7\hspace{-3mm} &
  \hspace{-3mm}24.3\hspace{-3mm} & \hspace{-3mm}27.32\hspace{-3mm} & \hspace{-3mm}30.52\hspace{-3mm} \\ \hline
\end{tabular}
\caption{Fusion probabilities calculation results for reaction $\alpha + ^{44}{\rm Ca}$ for parametrization~\cite{Denisov.2005.PHRVA} by the minimization method, where the MIR approach is used in calculations of the penetrabilities.
$p_{0}$ \ldots $p_{16}$ are the fusion probabilities,
$\varepsilon_{1}$, $\varepsilon_{2}$ and $\varepsilon_{3}$ are functions of errors defined via the formula~(\ref{eq.2.3.11}),
$E_{\rm min}$ is the minimum of the potential well before the barrier (MeV),
$E_{\rm max}$ is the height of the potential barrier (MeV), and
$L_{\rm max}$ is the maximal value of the orbital momentum, used in the summation in eq.~(\ref{eq.2.1.1}).}
\label{table.app.2.1}
\end{table}
From such results, one can see that at a smaller maximal momentum $L_{\rm max}$ all probabilities are equal to unity. However, at increasing $L_{\rm max}$ starting from $L_{\rm max}=9$, the fusion probabilities at the first angular momentum values are close to zero.
This tendency of the vicinity to zero of the first fusion probabilities is present at increasing $L_{\rm max}$ up to the maximum of its possible value.
Such a tendency is also present for another reaction $\alpha + ^{40}{\rm Ca}$, after finding its own parametrization of the $\alpha$-nucleus potential (see Tabls.~\ref{table.3.5.4} and \ref{table.3.5.5}).
Such a picture clearly demonstrates the failure of the idea for a sharp angular momentum cutoff approach introduced by Glas and Mosel~\cite{Glas.1975.NPA, Glas.1974.PRC}.
A more accurate description of the tendencies of the fusion probabilities at smaller and larger angular momenta allows us to describe the difference between the experimental data~\cite{Eberhard.1979.PRL} for reactions $\alpha + ^{40}{\rm Ca}$ and $\alpha + ^{44}{\rm Ca}$.
This indicates the importance of cancelation of the sharp angular momentum cutoff approach (which has produced countless papers) in the capture and synthesis tasks if we want to study the fusion processes.


\section{Calculations of the potential parameters by the minimization method
\label{sec.app.3}}

To estimate errors given by the different methods for the capture of $\alpha + ^{40}{\rm Ca}$ and $\alpha + ^{44}{\rm Ca}$, such calculations should be compared for the same parametrization.
We choose the parametrization from~\cite{Denisov.2005.PHRVA}, as this paper was directly oriented to the study of $\alpha$-capture processes, and the
authors provide the parameters for such reactions.
We include into the analysis our MIR approach, the WKB approach and the approaches developed by Hill and Wheeler in \cite{Hill_Wheeler.1953.PR} and Wong in~\cite{Wong.1973.PRL} to calculate the penetrability for the inverse parabolic barrier, which approximates the original potential barrier (we shall call them \emph{Hill-Wheeler and Wong's approaches}).
From the results presented in Tabl.~\ref{table.3.5.1}, one can see that the MIR method has half as many errors as the WKB calculations,
while the Hill-Wheeler and Wong's formulas give estimations of errors that are similar to those of
the MIR approach.
\begin{table}
\begin{center}
\begin{tabular}{|c|c|c|c|c|c|c|} \hline
 Method & \multicolumn{3}{|c|}{$\alpha + ^{40}{\rm Ca}$} &
 \multicolumn{3}{|c|}{$\alpha + ^{44}{\rm Ca}$} \\
 \cline{2-7}
               & $\varepsilon_{1}$ & $\varepsilon_{2}$ & $\varepsilon_{3}$ & $\varepsilon_{1}$ & $\varepsilon_{2}$ & $\varepsilon_{3}$ \\ \hline
  MIR method        & 0.5154 & 0.3873 & 0.0687 & 0.3881 & 0.3238 & 0.0561 \\
  WKB approach      & 0.9233 & 0.5251 & 0.1030 & 0.8401 & 0.4936 & 0.0957 \\
  Hill-Wheeler app. & 1.3923 & 0.6236 & 0.1306 & 0.3927 & 0.3265 & 0.0569 \\
  Wong formula      & 0.6834 & 0.4527 & 0.0829 & 0.5507 & 0.4007 & 0.0713\\ \hline
\end{tabular}
\end{center}
\caption{Errors obtained by different approaches for cross-sections of the capture reactions $\alpha + ^{40}{\rm Ca}$ and $\alpha + ^{44}{\rm Ca}$ with the potential parametrization given in~\cite{Denisov.2005.PHRVA},
where we do not include calculations of the fusion probabilities.
$\varepsilon_{1}$, $\varepsilon_{2}$ and $\varepsilon_{3}$ are functions of the errors defined by the formula~(\ref{eq.2.3.11}).
From the presented results, one can see that for the same parametrization~\cite{Denisov.2005.PHRVA}, the MIR method decreases errors twice in comparison with the WKB calculations, while the Hill-Wheeler and Wong's formulas give estimations of errors that are similar to those of the MIR approach.
}
\label{table.3.5.1}
\end{table}
From the first consideration, it strangely was not clear why the difference between the results obtained by the MIR approach and Hill-Wheeler and Wong's formulas was so small. The MIR approach essentially includes more formalism (and computer resources) for the approximated potential barrier and calculations of its penetrability than the Hill-Wheeler and Wong's formulas.
However, from such a point, it has become clear that the \emph{errors are only the compared difference between the calculations and the experimental data of the cross-sections, and they absolutely ignore the level of accuracy of the approximation of the original barrier}. According to such logic, we can take another potential, calculate its penetrability by any approximated formula, and, if this new value is closer to the experimental data, conclude the success of the used approach. However, this is absolutely the wrong conclusion. The error should therefore also take into account the level of the approximation of the original barrier by the method used.
As the WKB approach has worse accuracy, we do not include it in any further analysis.

Let us see which parameters of the $\alpha$-nucleus potential the studied methods give. The results obtained by the optimization technique (we choose the Gradient descent method) for $\alpha +\, {^{44}\!{\rm Ca}}$ are presented in Tabl.~\ref{table.3.5.2}, where we do not include calculations of the fusion probabilities. One can see that the estimations for the radius, $r_{\rm m}$, are very close for any used approach with parametrization~\cite{Denisov.2005.PHRVA}.
The strength of the potential, $V_{0}$, obtained by the MIR approach and Wong's formula are close to parametrization~\cite{Denisov.2005.PHRVA}, in contrast to the Hill-Wheeler approach. All three approaches essentially give different estimations for diffuseness, $d$, in comparison with the results from~\cite{Denisov.2005.PHRVA}.
It now becomes clear that the included formalism with different approximations of the original barrier gave different estimations of the parameters at similarly obtained errors.
This result underlines the importance of including the level of the approximation of the initial barrier when determining the error of the method.

\begin{table}
\begin{center}
\begin{tabular}{|c|c|c|c|c|c|c|c|c|} \hline
 & \multicolumn{5}{|c|}{Parameters of potential} &
 \multicolumn{3}{|c|}{errors}
  \\
 \cline{2-9}
       & $V_{0}$, Mev & $r_{\rm m}$, fm & $d$, fm &  $E_{\rm min}$ & $E_{\rm max}$ & $\varepsilon_{1}$ & $\varepsilon_{2}$ & $\varepsilon_{3}$ \\ \hline
  MIRM & 34.091 &    6.6855   & 0.1789 & -24.04 & 7.48 & 0.0738 & 0.0855 & 0.0129 \\
  HWA  & 16.463 &    6.6866   & 0.1568 &  -6.67 & 7.71 & 0.0761 & 0.0987 & 0.0151 \\
  WoF  & 33.155 &    6.6864   & 0.1455 & -23.30 & 7.66 & 0.0730 & 0.0879 & 0.0129 \\ \hline
\end{tabular}
\end{center}
\caption{Parameters of the potential obtained via minimization of the function of the error $\varepsilon_{1}$ for the capture reaction $\alpha + ^{44}{\rm Ca}$, where we do not include calculations of the fusion probabilities.
$\varepsilon_{1}$, $\varepsilon_{2}$ and $\varepsilon_{3}$ are error functions defined by formulas ~(\ref{eq.2.3.11}),
$E_{\rm min}$ is the minimum of the potential well before the barrier (MeV), and
$E_{\rm max}$ is the height of the potential barrier (MeV).
MIRM, HWA, and WoF are the MIR method, Hill-Wheeler approach, and Wong's formula, respectively.
One can see that after the minimization procedure,
the errors are less than a quarter of those in
the previous calculations in Tabl.~\ref{table.3.5.1}.
However, at a similar accuracy of experimental data description, different methods give essentially differences between parameters $V_{0}$ and $d$.
From here, it follows that the level of agreement with the experimental data on the basis of the given errors is not sufficient for extracting reliable information about the parameters of the interacting potential.
}
\label{table.3.5.2}
\end{table}

\section{Comparison of the approach of Hill and Wheeler and Wong's formula vs. the MIR and WKB approaches
\label{sec.app.4}}

\begin{figure}[htbp]
\centerline{%
\includegraphics[width=78mm]{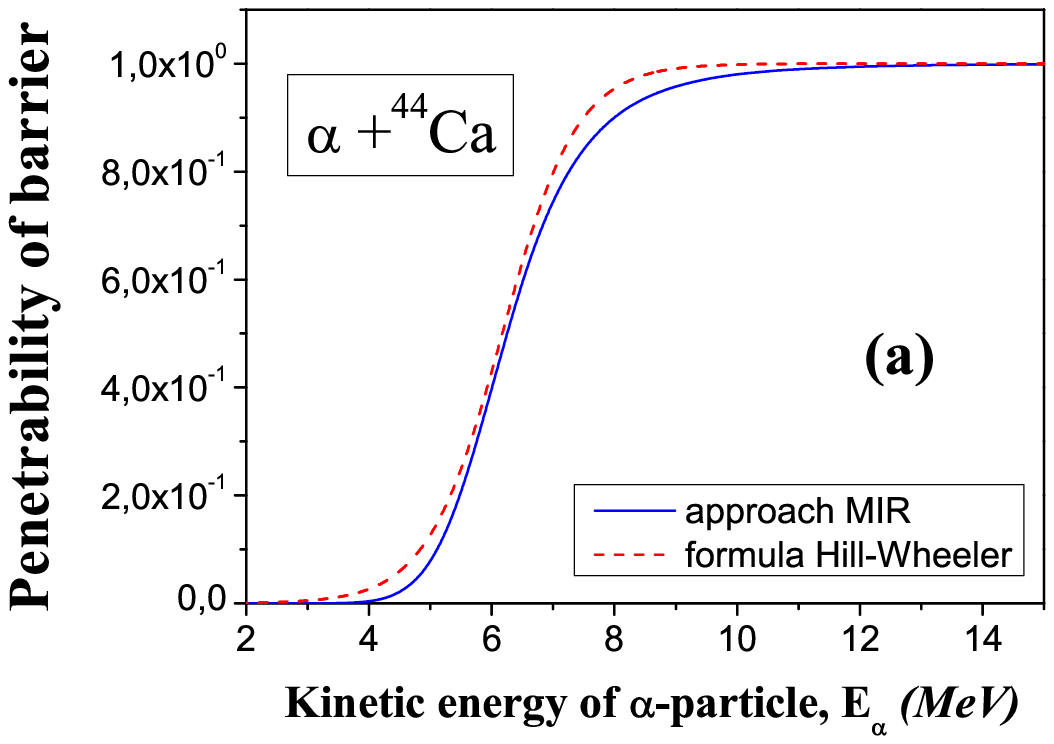}
\hspace{-6mm}\includegraphics[width=78mm]{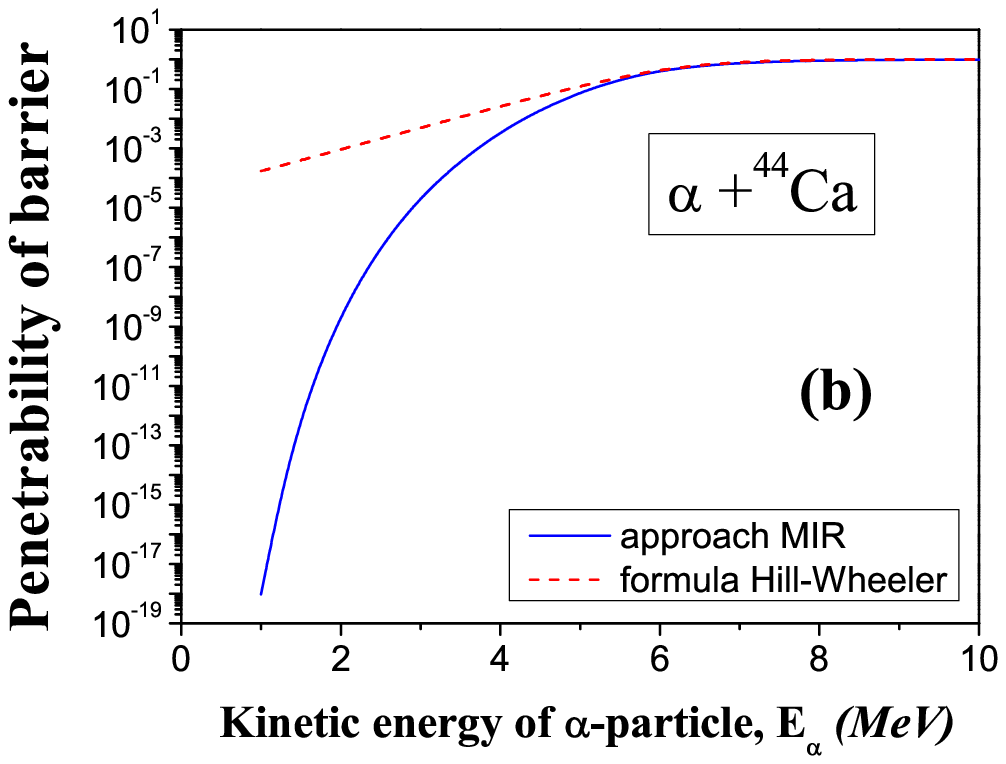}}
\vspace{-5mm}
\caption{\small 
The penetrabilities of the barrier for the reaction $\alpha + ^{44}{\rm Ca}$ at $l=0$ calculated by the MIR approach (solid blue line) and by the Hill-Wheeler formula (dashed red line) (parameters of calculations: 1000 intervals at $r_{\rm max}=70$~fm).
On the linear scale, we obtain a sufficiently close agreement between the penetrabilities calculated by the two approaches (see figure (a)).
However, a more accurate study of the logarithmic scale shows a higher estimation for the penetrability obtained by the Hill-Wheeler formula (see figure~(b)).
Now, it becomes clear that the application of such a formula for the determination of penetrabilities in the nuclear decay tasks can lead to essential errors in estimations of half-lives.
In fusion tasks, the Hill-Wheeler formula is saved by the fact that for the determination of the capture cross-section, we need to summarize up to 18 curves of the penetrabilities with shifted maxima
(only where this formula is working).
\label{fig.3.5.2}}
\end{figure}

\begin{figure}[htbp]
\centerline{%
\includegraphics[width=78mm]{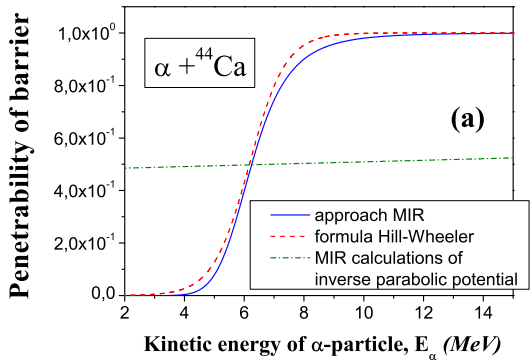}}
\vspace{-5mm}
\caption{\small 
The penetrability of the inverse parabolic barrier used in the approach of Hill and Wheeler in~\cite{Hill_Wheeler.1953.PR} at $l=0$ for reaction $\alpha + ^{44}{\rm Ca}$ as calculated by the MIR method (parameters of calculations: 1000 intervals at $r_{\rm max}=70$~fm).
Such calculations by the MIR approach are presented by the dash-dotted green line.
For comparative analysis, we add calculations of the penetrability of the original barrier by the MIR approach (see solid blue line) and by the Hill-Wheeler formula (see dashed red line), shown in the previous figure.
This result clearly shows that the MIR method essentially gives a different penetrability than the Hill-Wheeler formula.
The reason for such a difference between the calculations obtained by the two approaches will become clearer if we compare the
shapes of the inverse parabolic barrier with that of the original barrier.
Such a comparison shows that there are no similarities, with the exception of a small neighborhood near the barrier maximum (which gives close estimations of the cross-sections).
\label{fig.3.5.3}}
\end{figure}

The direct comparison of the calculations from the MIR approach and the Hill-Wheeler formula shows that the penetrability obtained by the Hill and Wheeler approach is essentially higher at smaller energies (see Fig.~\ref{fig.3.5.2}),
but this overestimated result has no visible influence on the cross-section.
However, this result is not correct in a fully quantum consideration, where we essentially have an influence of the shape of the barrier on the estimation of the penetrability.
To clarify the role of the shape of the inverse parabolic barrier (used in the Hill-Wheeler approach), we have calculated the penetrability of such a barrier by the MIR method.
The result of such calculations is presented in Fig.~\ref{fig.3.5.3} (see dash-dotted green line in this figure), which is strongly different from the estimation of the Hill-Wheeler formula
(see dashed red line in this figure).
Only the penetrability at the point of intersection between all the curves in such a figure plays a role in the calculations of the cross-section, which explains the not bad agreement of the calculations by the approach of Hill and Wheeler with the experimental data. However, from the analysis and calculations above, it is clear that good agreement of the calculations
by the approach of Hill and Wheeler with the experimental data does not confirm the validity of such a formula.
In Tabl.~\ref{table.3.5.3}, we demonstrate the convergence of calculations of the penetrability by the MIR approach
for such an inverse parabolic barrier.
\begin{table}
\begin{center}
\begin{tabular}{|c|c|c|c|c|c|c|} \hline
 Number of & \multicolumn{3}{|c|}{$r_{\rm max} = 50$~fm} & \multicolumn{3}{|c|}{$r_{\rm max} = 200$~fm} \\
 \cline{2-7}
  intervals & 1 MeV & 15 MeV & 30 MeV &  1 MeV  &  15 MeV &  30 MeV \\ \hline
  10000 & 0.48634 & 0.53097 & 0.57703 & 0.49473 & 0.51760 & 0.54100 \\
  20000 & 0.48086 & 0.52368 & 0.56866 & 0.44806 & 0.49275 & 0.54366 \\
  30000 & 0.48226 & 0.52432 & 0.56828 & 0.47961 & 0.51829 & 0.55901 \\
  40000 & 0.48018 & 0.52204 & 0.56617 & 0.48752 & 0.53010 & 0.57372 \\
  50000 & 0.48019 & 0.52197 & 0.56601 & 0.47070 & 0.51346 & 0.55993 \\ \hline
  $V (r_{\rm max})$ & \multicolumn{3}{|c|}{-3747000} & \multicolumn{3}{|c|}{-95000000} \\ \hline
\end{tabular}
 \end{center}
\caption{
The penetrability of the inverse parabolic barrier for $\alpha + ^{44}{\rm Ca}$ at $l=0$ as calculated by the MIR method for different values of the following calculated parameters:
The number of intervals and external boundary $r_{\rm max}$.
The parameters of the inverse parabolic barrier are determined according to the Hill-Wheeler approach~\cite{Hill_Wheeler.1953.PR}, which approximates the original potential barrier with parametrization~\cite{Denisov.2005.PHRVA}.
In such calculations we do not include the fusion probabilities.
$V (r_{\rm max})$ is the inverse parabolic potential at point $r_{\rm max}$ (MeV).
}
\label{table.3.5.3}
\end{table}
Moreover, the WKB approach includes the shape of the original barrier in the calculations. It is therefore more correct and accurate in the estimations of the penetrabilities of the barriers in fusion tasks (in comparison with Hill-Wheeler approach and Wong's formula).


\section{Comparison of the MIR approach vs. the Numerov approach
\label{sec.app.5}}

The penetrability and reflection coefficients can be determined on the basis of wave functions obtained via the direct integration of the radial Schr\"{o}dinger equation with high accuracy using the Numerov technique.
In this section, we shall compare the effectiveness of such a method in determining the studied coefficients in comparison with the MIR approach.
For simplicity of analysis, we shall study the reaction $\alpha + ^{44}{\rm Ca}$ for case $l=0$.

We use the same separation of the radial region on $N$ regions as defined in (\ref{eq.2.2.1}), and we assume that the capture of the $\alpha$-particle with the nucleus occurs in the first region after its tunneling through the barrier.
We define the normalized radial wave function as
\begin{equation}
\chi (r) =
\left\{
\begin{array}{lll}
\vspace{1mm}
  \exp (- ik_{1} r), & \mbox{\rm at } r_{\rm min} < r \le r_{1} & \mbox{\rm (region 1)}, \\
  A_{2}\, c_{2} (r) + B_{2}\, d_{2}(r), & \mbox{\rm at } r_{1} \le r \le r_{\rm max},
\end{array}
\right.
\label{eq.app.5.1}
\end{equation}
where $A_{2}$ and $B_{2}$ are unknown complex amplitudes, and
$c_{2}(r)$ and $d_{2}(r)$ are linear independent partial solutions of the radial Schr\"{o}dinger equation with potential defined by (\ref{eq.2.3.1})--(\ref{eq.2.3.4}). These amplitudes should be found from continuity conditions of the full wave function and its derivative at point $r_{1}$.
As a result, we obtain:
\begin{equation}
\begin{array}{lll}
\vspace{1mm}
  \Re (A_{2}) =
    N_{0}\; \bigl[\cos (k_{1}r_{1})\: d_{2}^{\prime}(r_{1}) + k_{1} \sin (k_{1}r_{1})\: d_{2}(r_{1}) \bigr], \\

\vspace{1mm}
  \Im (A_{2}) =
    N_{0}\; \bigl[k_{1} \cos (k_{1}r_{1})\: d_{2}(r_{1}) - \sin (k_{1}r_{1})\: d_{2}^{\prime}(r_{1}) \bigr], \\

\vspace{1mm}
  \Re (B_{2}) =
    -\, N_{0}\; \bigl[\cos (k_{1}r_{1})\: c_{2}^{\prime}(r_{1}) + k_{1} \sin (k_{1}r_{1})\: c_{2}(r_{1}) \bigr], \\

  \Im (B_{2}) =
    -\, N_{0}\; \bigl[k_{1} \cos (k_{1}r_{1})\: c_{2}(r_{1}) - \sin (k_{1}r_{1})\: c_{2}^{\prime}(r_{1}) \bigr],
\end{array}
\label{eq.app.5.2}
\end{equation}
where
\begin{equation}
  N_{0} =
  \displaystyle\frac{1}
    {c_{2}(r_{1})\, d_{2}^{\prime}(r_{1}) - c_{2}^{\prime}(r_{1})\, d_{2}(r_{1})}.
\label{eq.app.5.3}
\end{equation}
We define the partial wave functions $c_{2}(r)$ and $d_{2}(r)$ so
that in the asymptotic region, starting from some selected boundary $r_{\rm as}$,
they corresponds to Coulomb functions as
$ c_{2}(r) = G_{l=0}(r)$, $d_{2}(r) = F_{l=0}(r)$.
To determine the unknown wave functions and their derivatives in region $r_{1} \le r \le r_{\rm as}$ (where the nuclear component of the potential is important), we calculate them numerically using the Numerov approach.
One can rewrite the full wave function in the asymptotic region via a combination of
the incident and reflected waves relative to the barrier as
\begin{equation}
  \chi (r) =
  A_{\rm inc}\, \bigl[G_{0}(r) - i\,F_{0}(r) \bigr] +
  A_{\rm ref}\, \bigl[G_{0}(r) + i\,F_{0}(r) \bigr],
\label{eq.app.5.4}
\end{equation}
where we have introduced the amplitudes $A_{\rm inc}$ and $A_{\rm ref}$ for the incident wave and reflected wave.
Such amplitudes can be rewritten as
\begin{equation}
\begin{array}{lcl}
  A_{\rm inc} = \displaystyle\frac{A_{2} + i\,B_{2}}{2}, &
  A_{\rm ref} = \displaystyle\frac{A_{2} - i\,B_{2}}{2}
\end{array}
\label{eq.app.5.5}
\end{equation}
and we have the amplitude of transition equaling unity, $A_{\rm tr}=1$.
Taking into account the property
$ G_{0}\, F_{0}^{\prime} - G_{0}^{\prime}\, F_{0} = 1$
for the Coulomb functions,
we obtain the following solutions for the penetrability and reflection coefficients:
\begin{equation}
\begin{array}{lcl}
  T = \displaystyle\frac{4 k_{1}}{k} \displaystyle\frac{1}{f_{1} + f_{2}}, &
  R = \displaystyle\frac{f_{1} - f_{2}}{f_{1} + f_{2}},
\end{array}
\label{eq.app.5.4}
\end{equation}
where
\begin{equation}
\begin{array}{lcl}
\vspace{1mm}
  f_{1} = [\Re (A_{2})]^{2} + [\Im (A_{2})]^{2} + [\Re (B_{2})]^{2} + [\Im (B_{2})]^{2}, \\
  f_{2} = 2\, \bigl[ \Im (A_{2})\: \Re (B_{2}) - \Re (A_{2})\: \Im (B_{2}) \bigr].
\end{array}
\label{eq.app.5.5}
\end{equation}

To compare the effectiveness of the Numerov and MIR methods in determining the penetrability and reflection coefficients, we calculate these coefficients in dependence on the external boundary $r_{\rm max}$ and number of intervals $N$ by both approaches.
The results of these calculations are presented in Tables~\ref{table.app.5.1} and~\ref{table.app.5.2}.
\begin{table}
\begin{center}
\begin{tabular}{|c|c|c|c|c|c|} \hline
 $r_{\rm max}$, fm & \multicolumn{3}{|c|}{Numerov approach} & \multicolumn{2}{|c|}{MIR approach} \\
 \cline{2-6}
  & $T$ & $R$ & $\varepsilon_{1}$ &  $T$  &  $R$ \\ \hline
   50 & 0.890661 & 0.109292 & $4.5 \times 10^{-5}$ & 0.99997095707 & $2.90429 \times 10^{-5}$ \\
   75 & 0.890581 & 0.109371 & $4.6 \times 10^{-5}$ & 0.99997116245 & $2.88375 \times 10^{-5}$ \\
  100 & 0.890661 & 0.109292 & $4.5 \times 10^{-5}$ & 0.99997139532 & $2.86046 \times 10^{-5}$ \\
  150 & 0.890822 & 0.109133 & $4.4 \times 10^{-5}$ & 0.99997145272 & $2.85472 \times 10^{-5}$ \\
  200 & 0.891300 & 0.108659 & $3.9 \times 10^{-5}$ & 0.99997147408 & $2.85259 \times 10^{-5}$ \\
  250 & 0.890982 & 0.108974 & $4.2 \times 10^{-5}$ & 0.99997142915 & $2.85708 \times 10^{-5}$ \\ \hline
\end{tabular}
\end{center}
\caption{
The penetrability $T$ and reflection $R$ of the barrier with parametrization~\cite{Denisov.2005.PHRVA} for $\alpha + ^{44}{\rm Ca}$ at $l=0$ and energy $E=25$~MeV calculated by the Numerov approach and MIR method, for different values of the external boundary $r_{\rm max}$.
Here,
$\varepsilon_{1} = ||T + R| - 1|$ is the error calculated to test the obtained coefficients;
for all MIR calculations, we have $\varepsilon_{1} < 10^{-16}$,
we do not include the fusion probabilities, and the number of intervals is the same and equals 10000.
%
}
\label{table.app.5.1}
\end{table}
\begin{table}
\begin{center}
\begin{tabular}{|c|c|c|c|c|c|} \hline
 Number of & \multicolumn{3}{|c|}{Numerov approach} & \multicolumn{2}{|c|}{MIR approach} \\
 \cline{2-6}
  intervals & $T$ & $R$ & $\varepsilon_{1}$ &  $T$  &  $R$  \\ \hline 
    100 & 0.934547 & 0.065452 & $8.9 \times 10^{-8}$ & 0.95572409894 & 0.044275901 \\ 
   1000 & 0.894080 & 0.105897 & $2.1 \times 10^{-5}$ & 0.99997192337 & $2.80766 \times 10^{-5}$ \\ 
  10000 & 0.890822 & 0.109133 & $4.4 \times 10^{-5}$ & 0.99997145272 & $2.85472 \times 10^{-5}$ \\ 
  25000 & 0.890726 & 0.109228 & $4.5 \times 10^{-5}$ & 0.99997144515 & $2.85548 \times 10^{-5}$ \\ 
  50000 & 0.890533 & 0.109419 & $4.7 \times 10^{-5}$ & 0.99997144338 & $2.85566 \times 10^{-5}$ \\ \hline 
\end{tabular}
\end{center}
\caption{
The penetrability $T$ and reflection $R$ of the barrier with parametrization~\cite{Denisov.2005.PHRVA} for $\alpha + ^{44}{\rm Ca}$ at $l=0$ and energy $E=25$~MeV calculated by the Numerov approach and MIR method for different values of the number of intervals.
Here,
$\varepsilon_{1} = ||T + R| - 1|$ is the error used to test the calculated coefficients,
and for all MIR calculations, we obtain $\varepsilon_{1} < 10^{-16}$.
The external boundary is $r_{\rm max} = 150$~MeV, and
in such calculations we do not include the fusion probabilities.
}
\label{table.app.5.2}
\end{table}
According to such results, we conclude the following:
\begin{enumerate}
\item
To estimate the accuracy of the determination of the penetrability and reflection coefficients, we use the standard test of quantum mechanics, calculating the parameter $\varepsilon_{1} = ||T + R| - 1|$ by each method.
The Numerov approach determines these coefficients with accuracy $\varepsilon_{1} = 10^{-5}$, while the accuracy of the MIR approach is restricted by computer capability (in the Tables above, we provide the MIR calculations at $\varepsilon_{1} \le 10^{-16}$ for any chosen $r_{\rm max}$ and $N$).

\item
Analyzing the stability in determining the penetrability and reflection coefficients after varying the parameters $r_{\rm max}$ and $N$, we conclude that the MIR approach is essentially more accurate than the Numerov approach.
The accuracy of the determination of these coefficients by the Numerov approach is restricted by the limit $10^{-3}$: we have only 3 stable digits for these coefficients at the variation of $r_{\rm max}$ and number of intervals $N$.
At the same time, the accuracy of the determination of these coefficients by the MIR approach is infinitely increased by increasing the number of intervals $N$:
we reach the first 8 stable digits of these coefficients when $N \to 50000$
(see Tables~\ref{table.app.5.1} and~\ref{table.app.5.2}).
\end{enumerate}


\section{Role of the nuclear deformations in the $\alpha$-capture task
\label{sec.app.6}}

In this section, we shall analyze how the obtained results are changed if we take into account the nuclear deformation of the target-nucleus. In this paper, we shall consider the case of axial deformation of the nuclei.
For the $\alpha$-nucleus interactions, we use the potential with parametrization
defined in \cite{Denisov.2005.PHRVA} which was found for the $\alpha$-capture task with deformed nuclei.
The cross-section for $\alpha$-capture by nuclei with axial deformation can be defined as
\begin{equation}
  \sigma_{\rm capture} (E) =
  \displaystyle\frac{\pi\, \hbar^{2}}{2\,m\,\tilde{E}}\;
  \displaystyle\int\limits_{0}^{\pi / 2}
    \displaystyle\sum\limits_{l=0}^{+\infty}
    (2l+1)\, T_{l}(\theta)\, P_{l}\;
    \sin\theta\, d\theta,
\label{eq.app.6.1}
\end{equation}
where
$\theta$ is the angle between
the axial symmetry axis of the deformed nucleus and
the line connected centers-of-masses of the $\alpha$-particle and the nucleus,
$T_{l}(\theta)$ is the penetrability of the barrier, which depending on the angle $\theta$, and
$P_{l}$ is the probability for fusion of the $\alpha$-particle and nucleus.

\begin{figure}[htbp]
\centerline{%
\includegraphics[width=82mm]{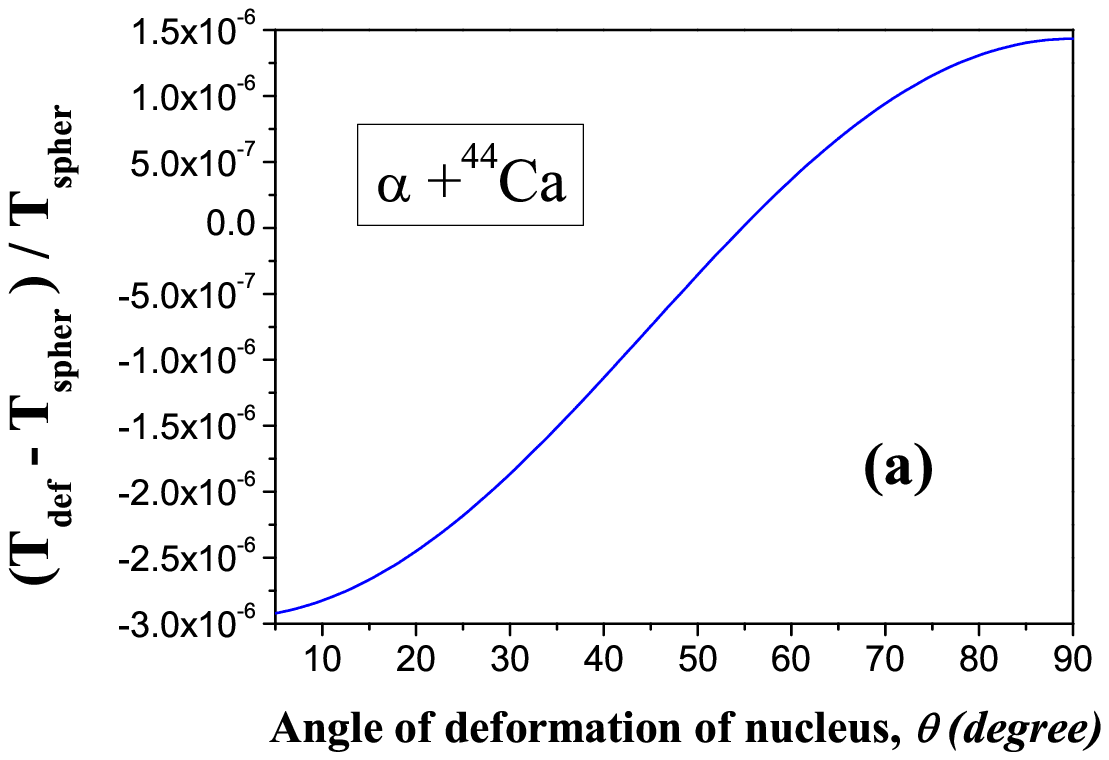}
\hspace{-6mm}\includegraphics[width=76mm]{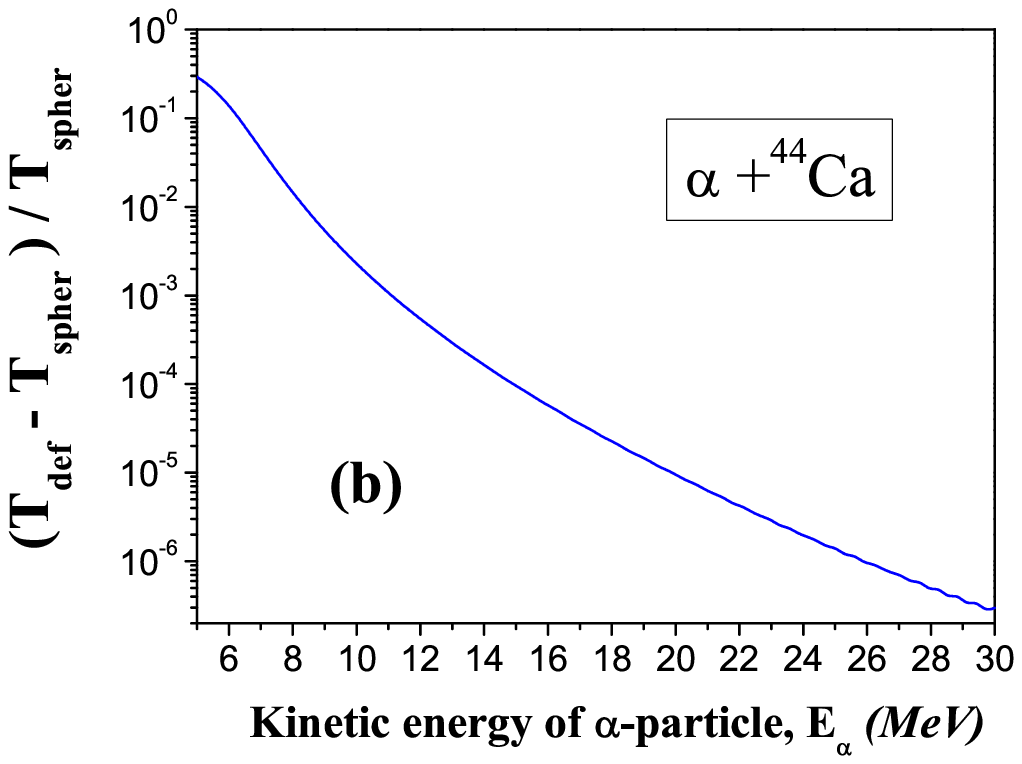}}
\vspace{-5mm}
\caption{\small 
The ratio of the penetrabilities of the barriers for the deformed nucleus vs. the spherical one, as a function of the angle of deformation $\theta$ at the incident energy of the $\alpha$-particle of $25$~MeV (a) and the energy of deformation $\theta$ at the angle of deformation of $\theta=90^{\circ}$ and (b)
for reaction $\alpha + ^{44}{\rm Ca}$ at $l=0$ calculated by the MIR method
(parameters of calculations: 10000 intervals at $r_{\rm max}=150$~fm).
Here,
$T_{\rm def}$ is the penetrability of the barrier for the deformed nucleus,
$T_{\rm spher}$ is the penetrability of the barrier for the spherical nucleus.
%
\label{fig.app.6.1}}
\end{figure}

For simplicity of the analysis, we shall study the capture reaction of $\alpha + ^{44}{\rm Ca}$.
The experimental data on the static quadrupole deformation parameter $\beta$ is taken from the RIPL-2 database \cite{ripl-2} (for $^{44}{\rm Ca}$ we have $\beta = 0.175423$).
First, we shall analyze how the penetrability of the barrier is changed after the inclusion of the deformation of the nucleus.
In Fig.~\ref{fig.app.6.1}~(a), we present our calculations for the penetrability, as a function of
the angle of deformation $\theta$ for case $l=0$.
We see that the largest influence of the nuclear deformation exists for angles of $0^{\circ}$ and $90^{\circ}$, while at $\theta = 55^{\circ}$, such an influence is missing.
Choosing the angle $\theta=90^{\circ}$, we can then analyze the role of nuclear deformation in the calculations of the penetrability, as a function of the energy of the incident $\alpha$-particle. The results of such calculations are presented in the subsequent Fig.~\ref{fig.app.6.1}~(b). From here, one can see that high importance of taking into account the nuclear deformations for the lowest energy of the $\alpha$-particle
(for $E_{\alpha} = 5.1$~MeV, where we obtain a ratio of $(T_{\rm def} - T_{\rm spher}) / T_{\rm spher} = 0.27739$).

\begin{figure}[htbp]
\centerline{\includegraphics[width=155mm]{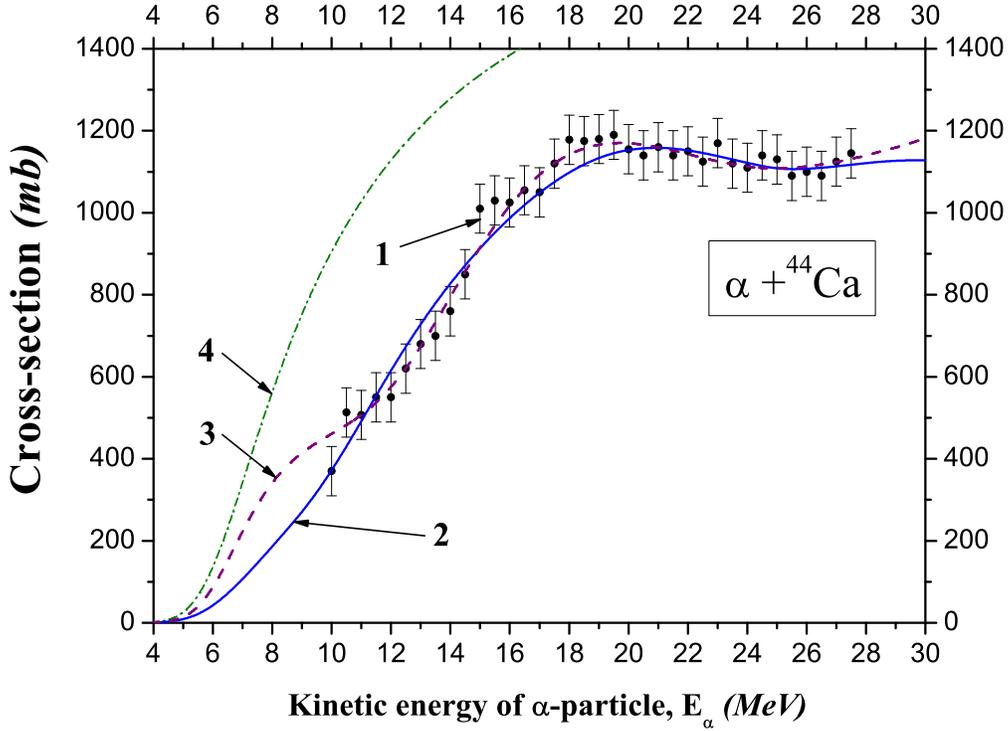}}
\vspace{-9mm}
\caption{\small 
The capture cross-sections of the $\alpha$-particle by the $^{44}{\rm Ca}$ nucleus obtained by the MIR method
(parameters of calculations: 10000 intervals at $r_{\rm max}=70$~fm, parametrization~\cite{Denisov.2005.PHRVA}).
Here, the data labeled 1 are the experimental data extracted from \cite{Eberhard.1979.PRL},
solid blue line 2 is the cross-section at $l_{\rm max}=17$ for the spherical nucleus,
dash-dotted purple line 3 is the cross-section at $l_{\rm max}=17$ for the deformed nucleus, and
dash dotted green line 4 is the cross-section at $l_{\rm max}=15$ for the spherical nucleus
(cross-sections are defined by (\ref{eq.app.6.1}), where $l_{\rm max}$ is the upper limit in the summation).
Lines 2 and 3 are obtained with the included fusion probabilities, line 4 without the fusion probabilities.
\label{fig.app.6.2}}
\end{figure}

In Tabl.~\ref{table.app.6.1}, we present the results of the determination of the fusion probabilities by the MIR approach for $\alpha + ^{44}{\rm Ca}$ after the inclusion of the deformed shape of the nucleus in the calculations. One can see that the MIR approach allows us to obtain such coefficients for the deformed nuclei with an
accuracy similar to that of the spherical nuclei.
\begin{table}
\begin{center}
\begin{tabular}{|c|c|c|c|c|c|c|c|c|c|c|} \hline
  Nucleus & $p_{0}$ & $p_{1}$ & $p_{2}$ & $p_{3}$ & $p_{4}$ & $p_{5}$ & $p_{6}$ & $p_{7}$ & $p_{8}$ & $p_{9}$ \\ \hline
  spherical & 0.04 & 0.03 & 0.01 & 0.01 & 0.43 & 0.31 & 0.62 & 0.31 & 0.71 & 1.00 \\ 
  deformed  & 0.14 & 0.48 & 0.62 & 0.32 & 0.56 & 0.14 & 0.22 & 0.18 & 0.16 & 1.00 \\ 
  \hline
\end{tabular}

\vspace{1mm}
\begin{tabular}{|c|c|c|c|c|c|c|c|c|c|} \hline
  Nucleus & $p_{10}$ & $p_{11}$ & $p_{12}$ & $p_{13}$ & $p_{14}$ & $p_{15}$ & $p_{16}$ & $p_{17}$ & $p_{18}$ \\ \hline
  spherical & 1.00 & 1.00 & 1.00 & 0.67 & 0.09 & 0.01 & 0.91 & 1.00 & 0.01 \\ 
  deformed  & 1.00 & 0.72 & 0.38 & 0.02 & 0.23 & 0.51 & 1.00 & 1.00 & 0.01 \\ 
  \hline
\end{tabular}
\end{center}
\caption{The fusion probabilities for the capture of the $\alpha$-particle by the spherical and deformed nuclei $^{44}{\rm Ca}$
obtained by the MIR approach
(parameters of calculations: 10000 intervals at $r_{\rm max}=70$~fm;
errors in obtaining results for the deformed nucleus are
$\varepsilon_{1}=0.0191$,
$\varepsilon_{2}=0.0263$ and
$\varepsilon_{3}=0.0040$).
}
\label{table.app.6.1}
\end{table}
The corresponding cross-section for the $\alpha$-capture by the deformed nucleus $^{44}{\rm Ca}$
with included fusion probabilities is given in Fig.~\ref{fig.app.6.2}.
One can see that line 3, obtained for the deformed nucleus describes the experimental data with good accuracy, similar to that of line 2 obtained for the spherical nucleus.
We observe the presence of small oscillations in the calculated spectrum after the inclusion of the nuclear deformation.




\end{document}